\global\def\draftcontrol{0}

   \def\versionno{ n2 thermal gap}

\catcode`\@=11

\expandafter\ifx\csname draftcontrol\endcsname\relax\global\def\draftcontrol{0}
\fi

{\count255=\time\divide\count255 by 60
\xdef\hourmin{\number\count255}
\multiply\count255 by-60\advance\count255 by\time
\xdef\hourmin{\hourmin:\ifnum\count255<10 0\fi\the\count255}}
\def\draftdate{\number\month/\number\day/\number\year\ \ \ \hourmin }

\newcommand\makepapertitle{\par
  \begingroup
    \renewcommand\thefootnote{\@fnsymbol\c@footnote}%
    \def\@makefnmark{\rlap{\@textsuperscript{\normalfont\@thefnmark}}}%
    \long\def\@makefntext##1{\parindent 1em\noindent
            \hb@xt@1.8em{%
                \hss\@textsuperscript{\normalfont\@thefnmark}}##1}%
     \newpage
     \global\@topnum\z@   
     \@makepapertitle
     \thispagestyle{empty}\@thanks
  \endgroup
  \setcounter{footnote}{0}%
  \global\let\thanks\relax
  \global\let\makepapertitle\relax
  \global\let\@makepapertitle\relax
  \global\let\@thanks\@empty
  \global\let\@author\@empty
  \global\let\@date\@empty
  \global\let\@title\@empty
  \global\let\title\relax
  \global\let\author\relax
  \global\let\date\relax
  \global\let\and\relax
  \def\version{\let\version\@version\@gobble}
}
\def\@makepapertitle{%
  \newpage
   \ifnum\draftcontrol=1 {}
   \version\versionno
   \vskip 3em%
   \else
   \hfill\hbox to 3cm {\parbox{4cm}{\@pubnum}\hss}%
   \vskip 3em%
   \fi
   \begin{center}%
   \let \footnote \thanks
     {\LARGE {\@title}}%
     \vskip 1.5em%
     {\normalsize
       \lineskip .5em%
       \begin{tabular}[t]{c}%
         \@author
       \end{tabular}\par}%
     \vskip 1.5em%
     {\@bstract}%
     \end{center}%
     \vskip 1.5em
     \@date%
   \par
}

\gdef\@pubnum{}
\def\pubnum#1{%
  \gdef\@pubnum{#1}}

\gdef\@bstract{}
\def\Abstract#1{%
  \gdef\@bstract{%
   \parbox{\textwidth-0pc}{%
   \centerline{\bf Abstract}\penalty1000%
\kern.2cm%
\noindent
\renewcommand\baselinestretch{1.0}%
{#1}}}
}

\def\ps@paper{\let\@mkboth\@gobbletwo%
     \ifnum\draftcontrol=1
    \def\@oddfoot{\hbox to \textwidth{\tiny \versionno \hfil\tiny\draftdate}%
    \hskip -\textwidth \hbox to \textwidth{\hfil\rm\thepage\hfil}}%
     \else\def\@oddfoot{\hbox to \textwidth{\hfil\rm\thepage\hfil}}
     \fi
     \let\@evenfoot\@oddfoot
}

\def\body{\clearpage
          \pagestyle{paper}
    }

\def\@version#1{\ifnum\draftcontrol=1
\typeout{}\typeout{#1}\typeout{}
\vskip3mm\centerline{\hbox{\fbox{\normalsize{\tt DRAFT -- #1 -- }
                   {\draftdate}}}}\vskip3mm
\fi}
\let\version\@version
\long\def\eqlabel#1{\ifnum\draftcontrol=1
                    \tag@false  
                    \tag*{(\theequation) \hbox to -0.2cm{\hspace{0cm}\small{#1}\hss}}
                    \refstepcounter{equation}
                    \edef\@currentlabel{\theequation}
                    \ltx@label{#1}          
                    \else
                    \label{#1}
                    \fi
                    }
\let\st@bibitem\@bibitem
\let\st@lbibitem\@lbibitem
\ifnum\draftcontrol=1
  \def\@bibitem#1{%
    \st@bibitem{#1}\a@@label{#1}\ignorespaces}
  \def\@lbibitem[#1]#2{%
    \st@lbibitem[#1]{#2}\a@@label{#2}\ignorespaces}
  \def\a@@label#1{%
    \gdef\a@lab{\smash{\normalfont\small#1}}
    \ifvmode
      \if@inlabel
        \global\setbox\@labels\hbox{%
          \llap{\a@lab\let\a@lab\relax
                \kern\@totalleftmargin\kern\marginparsep}%
          \box\@labels}%
      \fi
    \fi}
\fi

\documentclass[12pt,letterpaper]{article}

\usepackage{amsmath,amssymb,array,calc,epsfig,rotating,bm}
\usepackage[sort]{cite}
\usepackage{graphicx}
\usepackage{psfrag,verbatim}

\ifnum\draftcontrol=1
\fi

\renewcommand\baselinestretch{1.25}
\renewcommand\section{\@startsection {section}{1}{\z@}%
                                   {-3.5ex \@plus -1ex \@minus -.2ex}%
                                   {2.3ex \@plus.2ex}%
                                   {\normalfont\large\bfseries}}
\renewcommand\subsection{\@startsection{subsection}{2}{\z@}%
                                   {-3.25ex\@plus -1ex \@minus -.2ex}%
                                   {1.5ex \@plus .2ex}%
                                   {\normalfont\normalsize\bfseries}}
\renewcommand\subsubsection{\@startsection{subsubsection}{3}{\z@}%
                                   {-3.25ex\@plus -1ex \@minus -.2ex}%
                                   {1.5ex \@plus .2ex}%
                                   {\normalfont\normalsize\it}}
\renewcommand\paragraph{\@startsection{paragraph}{4}{\z@}%
                                   {-3.25ex\@plus -1ex \@minus -.2ex}%
                                   {1.5ex \@plus .2ex}%
                                   {\normalfont\normalsize\bf}}

\numberwithin{equation}{section}

\def\revise#1       {\raisebox{-0em}{\rule{3pt}{1em}}%
                     \marginpar{\raisebox{.5em}{\vrule width3pt\
                     \vrule width0pt height 0pt depth0.5em
                     \hbox to 0cm{\hspace{0cm}{%
                     \parbox[t]{4em}{\raggedright\footnotesize{#1}}}\hss}}}}

\newcommand\nxt[1]  {\\\fnxt#1}
\newcommand{\ie}{{\it i.e.,}\ }

\def\calc         {{\cal C}}

\def\cale         {{\cal E}}
\def\calf         {{\cal F}}

\def\call         {{\cal L}}
\def\calm         {{\cal M}}
\def\caln         {{\cal N}}
\def\calo         {{\cal O}}
\def\calp         {{\cal P}}

\def\calv         {{\cal V}}

\def\complex      {{\mathbb C}}

\def\del          {\partial}

\def\Im           {{\rm Im\hskip0.1em}}

\def\sqr#1#2{{\vcenter{\vbox{\hrule height.#2pt
 \hbox{\vrule width.#2pt height#1pt \kern#1pt
 \vrule width.#2pt}\hrule height.#2pt}}}}

\newcommand{\ft}[2]{{\textstyle{\frac{#1}{#2}}}}

\def\a{\alpha}
\def\b{\beta}
\def\w{\omega}
\def\r{\rho}
\def\dd{\delta}
\def\e{\epsilon}
\def\c{\chi}

\def\aa1{\phi}
\def\cc1{\psi}

\def\l{\lambda}

\def\Om{\Omega}

\def\hr{\hat{\rho}}

\def\ha{\hat{a}}

\def\lgb{\l_{GB}}
\def\hc{\hat{c}}
\def\ha{\hat{a}}
\def\hd{\hat{d}}

\catcode`\@=12

\begin{document}


\title{\bf  Black hole spectra in holography: consequences for equilibration of dual gauge theories}

\date{May 6, 2015}

\author{
Alex Buchel\\[0.4cm]
\it Department of Applied Mathematics\\
\it University of Western Ontario\\
\it London, Ontario N6A 5B7, Canada\\
\it Perimeter Institute for Theoretical Physics\\
\it Waterloo, Ontario N2J 2W9, Canada
}

\Abstract{
For a closed system to equilibrate from a given initial condition
there must exist an equilibrium state with the energy equal to the
initial one.  Equilibrium states of a strongly coupled gauge theory
with a gravitational holographic dual are represented by black holes.
We study the spectrum of black holes in Pilch-Warner geometry.  These
black holes are holographically dual to equilibrium states of strongly
coupled $SU(N)$ ${\cal N}=2^*$ gauge theory plasma on $S^3$ in the
planar limit. We find that there is no energy gap in the black hole
spectrum.  Thus, there is a priory no obstruction for equilibration of
arbitrary low-energy states in the theory via a small black hole
gravitational collapse.  The latter is contrasted with
phenomenological examples of holography with dual four-dimensional
CFTs having non-equal central charges in the stress-energy tensor
trace anomaly.
}

\makepapertitle

\body

\version\versionno
\tableofcontents

\section{Introduction and summary}\label{intro}
Consider an interacting system in a finite volume. Suppose that the theory is gapless ---
there are arbitrarily low-energy excitations. If a generic state in a theory equilibrates, 
there can not be a gap in the spectrum of equilibrium states in the theory. 
This obvious statement has a profound implication for strongly coupled 
gauge theories with an asymptotically AdS gravitational dual \cite{m1}. 
In a holographic dual the equilibrium states are realized by black holes
\cite{Aharony:1999ti}. Thus, if it is possible to prepare an arbitrary low-energy 
initial configurations in a holographic dual with a gapped spectrum of black holes, such 
states of the boundary gauge theory will never equilibrate. 
Correspondingly, the asymptotically AdS  dual  
is guaranteed to be stable against gravitational collapse for sufficiently 
small amplitude of the perturbations. Examples of this type would violate 
ergodicity from the field theory perspective.
 
In this paper we show that while it is possible to realize above scenario in a phenomenological 
(bottom-up) holographic example --- the Einstein-Gauss-Bonnet (EGB) gravity with a negative 
cosmological constant, it does not occur in a specific model of 
gauge theory/supergravity correspondence we consider --- the holographic duality
between $\caln=2^*$ $SU(N)$ gauge theory and the gravitational Pilch-Warner (PW) flow 
\cite{pw,bpp,cj}. 

From the gauge theory perspective, $SU(N)$ $\caln=2^*$  gauge theory is obtained from the parent $\caln=4$ SYM by giving a mass to 
$\caln=2$ hypermultiplet in the adjoint representation. 
In $R^{3,1}$ space-time, the low-energy effective action of the theory can be computed 
exactly \cite{Donagi:1995cf}. The theory has quantum Coulomb branch vacua $\calm_\calc$, parameterized 
by the expectation values of the complex scalar $\Phi$ in the $\caln=2$ vector multiplet, taking values in the Cartan subalgebra of the 
gauge group,
\begin{equation}
\Phi={\rm diag}(a_1,a_2,\cdots,a_N)\,,\qquad \sum_{i}a_i=0\,,
\eqlabel{vevs}
\end{equation}  
resulting in complex dimension of the moduli space
\begin{equation}
{\rm dim}_\complex\ \calm_\calc\ =\ N-1\,. 
\eqlabel{dim}
\end{equation}
In the large-$N$ limit, and for strong 't Hooft coupling, the holographic 
duality reduces to the correspondence between the gauge theory and 
type IIb supergravity. Since supergravities have finite number of light modes, 
one should not expect to see the full moduli space of vacua in $\caln=2$ 
examples of gauge/gravity correspondence. This is indeed what is happening:
the PW flow localizes on a semi-circle distribution of \eqref{vevs} with a
linear number density  \cite{bpp},
\begin{equation}
\begin{split}
&\Im(a_i)=0\,,\qquad a_i\in [-a_0,a_0]\,,\qquad a_0^2=\frac{m^2 g_{YM}^2 N}{4\pi^2}\,,\\
&\r(a)=\frac{8\pi}{m^2 g_{YM}^2}\ \sqrt{a_0^2-a^2}\,,\qquad \int_{-a_0}^{a_0}da\ 
\r(a)=N\,,
\end{split}
\eqlabel{pwdistr}
\end{equation}
where $m$ is the hypermultiplet mass. This holographic localization can be deduced entirely from the field theory 
perspective \cite{Buchel:2013id}, 
using the $S^4$-supersymmetric localization techniques \cite{Pestun:2007rz}. 
To summarize, $\caln=2^*$ holography is a well-understood nontrivial example of gauge/gravity correspondence 
that passes a number of highly nontrivial tests \cite{bpp,Buchel:2013id,Bobev:2013cja}.

We would like to compactify the background space of the $\caln=2^*$ strongly coupled 
gauge theory on $S^3$ of radius $\ell$ --- in a dual gravitational picture
we prescribe the boundary condition for the non-normalizable component of the 
metric in PW effective action to be that of $R\times S^3$. This is in addition to 
specifying  non-normalizable components (corresponding to $m$ in \eqref{pwdistr})
for the two PW scalars, dual to the mass deformation
operators of dimensions $\Delta=2$ and $\Delta=3$ of the gauge theory hypermultiplet mass term.
Thus, we produced  a holographic example of a strongly interacting 
system in a finite volume. The single dimensionless 
parameter\footnote{$\caln=2^*$ theory in Minkowski space-time has a scale associated 
with the Coulomb branch moduli distribution \eqref{pwdistr}. Once the theory is compactified 
on the $S^3$ the moduli space is lifted.}, so far, is $m \ell$. 
We proceed to construct  regular solutions of the PW effective 
gravitational action with the prescribes boundary condition, interpreting them 
as  vacua of $S^3$-compactified strongly coupled $\caln=2^*$ gauge theory.  
Using the standard holographic renormalization technique\footnote{For the model in hand 
this was developed in \cite{Buchel:2004hw}.} we compute the vacuum energy of the theory as 
a function of $m\ell$, $E_{vacuum}=E_{vacuum}(m\ell)$. We do not verify in this work whether 
described  $S^3$-compactifications preserve any supersymmetry; thus, it is important to
check the stability of the vacuum solutions. Previously, careful analysis of the
$S^4$-compactified PW holographic flows of \cite{Buchel:2013fpa} 
pointed to the discrepancy in the free energy of the 
solutions, compared with the localization prediction in \cite{Buchel:2013id}.
This discrepancy was resolved by identifying a larger  truncation 
\cite{Bobev:2013cja} (BEFP)\footnote{Of course, 
BEFP can itself be consistently truncated to PW.},   
where it was pointed out that preservation of the $S^4$-supersymmetry 
necessitates turning on additional bulk scalar fields. 
Stability of the PW embedding inside BEFP was discussed in  
\cite{Balasubramanian:2013esa}. We verify here that $S^3$-compactified 
PW vacua are stable within BEFP truncation. 
Having constructed vacuum solutions, we move to the discussion of the black hole spectrum.
We construct regular Schwarzschild black hole solutions in PW effective action, and compute 
$\dd E\equiv \dd E(m\ell, \ell_{BH}/L)\equiv E-E_{vacuum}(m\ell)$.   
We argue that there is no obstruction of initializing arbitrary low-energy 
excitations over the vacuum. Thus, one would expect no gap in the energy spectrum 
of PW black hole solutions, realizing equilibrium configurations of the strongly coupled 
$\caln=2^*$ gauge theory in the planar limit. Indeed, we find strong numerical evidence 
that 
\begin{equation}
\lim_{\ell_{BH}/L\to 0}\ \frac{\dd E(m\ell, \ell_{BH}/L)}{E_{vacuum}(m\ell=0)}\ =\ 0 \,.
\eqlabel{result}
\end{equation}

The rest of the paper is organized as follows. 
In the next section we discuss the spectrum of black holes in five-dimensional 
EGB gravity with a negative cosmological constant. These gravitational backgrounds 
can be interpreted as holographic duals to equilibrium states of strongly coupled 
conformal gauge theories with non-equal central charges in the stress-energy tensor trace
anomaly. We show that there is a gap in the spectrum of black holes. 
However, as one imposes constraints on EGB gravity coming from interpreting 
it as an effective description of gauge theory/string theory correspondence, 
the claim about the gap becomes unreliable --- higher derivative 
corrections, which are not under control, make order-one corrections to the
gap.  We follow up with the discussion in the $\caln=2^*$ holographic example. 
In the section \ref{action}
we review the PW effective action and its embedding 
within a larger BEFP truncation.  In section \ref{vacuum} we construct gravitational dual to vacuum 
states of $\caln=2^*$ gauge theory on $S^3$. Stability of the latter states within BEFP  truncation
is discussed in section \ref{vacuumstability}. In section \ref{bh} we study the spectrum of black holes in 
PW effective action.

\section{Black hole spectrum in Einstein-Gauss-Bonnet gravity}

Effective action of a five-dimensional Einstein-Gauss-Bonnet gravity with a negative cosmological constant 
takes form:
\begin{equation}
\begin{split}
S=&\frac{1}{2\ell_p^3}\int_{\calm_{5}}d^{5}z \sqrt{-g} 
\biggl(\frac{12}{L^2}+R+
\frac{\lgb}{2} L^2\left(R^2-4 R_{\mu\nu}R^{\mu\nu}
+R_{\mu\nu\r\sigma}R^{\mu\nu\rho\sigma}\right) 
\biggr)\,.
\end{split}
\eqlabel{eq:aisnotc}
\end{equation}
When interpreted in a framework of gauge theory/gravity correspondence\footnote{See \cite{Banerjee:2014oaa}
for a recent review.}, EGB action \eqref{eq:aisnotc} 
represents a holographic dual to a putative strongly coupled conformal theory with non equal central charges, 
$c\ne a$, of the boundary stress-energy tensor,
\begin{equation}
\begin{split}
&\langle T^\mu{}_\mu\rangle_{\rm CFT} =\frac{c}{16\pi^2}
I_4-\frac{a}{16\pi^2} E_4\,,\\
&E_4= r_{\mu\nu\rho\lambda}r^{\mu\nu\rho\lambda}-4
r_{\mu\nu}r^{\mu\nu}+r^2 \,,\\
& I_4=
r_{\mu\nu\rho\lambda}r^{\mu\nu\rho\lambda}-2  r_{\mu\nu}r^{\mu\nu}
+\frac 13r^2\,,
\end{split}
\eqlabel{eq:anomaly}
\end{equation}
where $E_4$ and $I_4$ are the four-dimensional Euler density and the square of the Weyl curvature
of the CFT background space-time.  The precise identification of the central charges is as follows:
\begin{equation}
\begin{split}
&c=\frac{\pi^2 \tilde{L}^3}{\ell_p^3}\left(1-2\frac{\lgb}{\b^2}\right)\,,\qquad 
a=\frac{\pi^2 \tilde{L}^3}{\ell_p^3}\left(1-6\frac{\lgb}{\b^2}\right)\,,\\
&\tilde{L}\equiv \b L\,,\qquad \b^2\equiv \frac 12 +\frac 12 \sqrt{1-4\lgb}\,.
\end{split}
\eqlabel{eq:defca}
\end{equation}
The gravitational dual to the vacuum state of a CFT on a three-sphere $S^3$ is a global $AdS_5$,
 \begin{equation}
ds^2 = \frac{L^2\b^2}{\cos^2 x} \left(
                                     - dt^2
                                     +{dx^2}
                                     +\sin^2x \, d\Omega^2_{3}
                                        \right) \, , \qquad x\in[0,\pi/2]\,,
\eqlabel{eq:adsmetric}
\end{equation}
where $d\Omega_3^2$ is the metric of $S^3$.  
Notice that $\lgb$ is restricted to be
\begin{equation}
\lgb\le \frac 14\,;
\eqlabel{lgbconst1}
\end{equation}
otherwise, there is simply no asymptotic AdS solution.
Following holographic renormalization 
of EGB gravity developed in \cite{Liu:2008zf,Banerjee:2014oaa}, we find that the 
vacuum energy (the mass) of \eqref{eq:adsmetric}, or  the 
Casimir energy from the boundary CFT perspective, is 
\begin{equation}
E_{vacuum}=\frac{3a}{4\tilde{L}}\,.
\eqlabel{eq:casimir}
\end{equation} 
 
Black holes (equilibrium configurations of EGB CFT) are found as a regular horizon solutions
within the metric ansatz,  
\begin{equation} 
ds^2 = \frac{L^2\b^2}{\cos^2 x} \left(
                                     -A(x) dt^2
                                     +\frac{dx^2}{A(x)}
                                     +\sin^2x \, d\Omega^2_{3}
                                        \right) \,.
\eqlabel{gbbh}
\end{equation}
The most general solution of equations of motion obtained from \eqref{eq:aisnotc} determine 
$A(x)$ is terms of a single parameter $M>0$,
\begin{equation}
\begin{split}
A=&~1-\frac{1}{2 \lgb} \biggl((2 \lgb-\b^2) \sin^2 x
+\biggl(4 \lgb (\b^2-2 \lgb) M \cos^4 x\\
&+(2 \lgb-\b^2)^2 \cos^4 x
-\b^4 (1-4 \lgb) \cos(2 x)\biggr)^{1/2}\biggr)\,.\\
\end{split}
\eqlabel{eq:statbh}
\end{equation} 
Furthermore, using the machinery of the holographic renormalization, the energy of the boundary CFT is 
\begin{equation}
E=\frac{3c}{4L \b} \biggl(\frac{\b^2-6\lgb}{\b^2-2\lgb}+4 M\biggr)
=\frac{3c}{4\tilde{L}} \biggl(\frac{a}{c}+4 M\biggr)\,.
\eqlabel{eq:energyxi}
\end{equation}
It is remarkable that the regular  Schwarzschild horizon in the geometry 
\eqref{gbbh}, \eqref{eq:statbh}  exists only provided \cite{Cai:2001dz,Buchel:2014dba}
\begin{equation}
M \ge 
\begin{cases}
\frac{1-\b^2}{2\b^2-1}\,, &{\rm if}\ \lgb>0\,,\\
{(\b^2-1)}{(2\b^2-1)}\,, &{\rm if}\ \lgb<0\,.
\end{cases}
\eqlabel{eq:defm}
\end{equation}
For positive $\lgb$, the bound comes requiring that $S^3$ remains finite
at the location of the horizon (otherwise the curvature at the horizon diverges).
For negative $\lgb$, violating the bound would render geometry complex (expression 
inside the square root in \eqref{eq:statbh} would turn negative 
for some $x\in (0,\pi/2)$).   

Constraints \eqref{eq:defm} imply the gap in $\dd E\equiv E-E_{vacuum}$ in the spectrum 
of EGB black holes,
\begin{equation}
\frac{\dd E}{|E_{vacuum}|}\ \ge\ \e_{gap}=\frac{4(1-\b^2)}{|6\b^2-5|}\times 
\begin{cases} 1\,,\qquad \lgb>0\,,\\
-(2\b^2-1)^2\,,\qquad \lgb<0\,,
\end{cases} 
\eqlabel{gbgap}
\end{equation}
With the only restriction \eqref{lgbconst1} on $\lgb$, 
$\e_{gap}$ is unbounded as $\lgb\to -\infty$ and $\lgb\to 5/36$.

We argue now that attempts to interpret EGB holography as an effective description 
of some gauge theory/string theory correspondence make the gap claim \eqref{gbgap} 
unreliable.   First,  causality of the holographic GB hydrodynamics requires that \cite{Buchel:2009tt}
\begin{equation}
-\frac{7}{36}\le\lgb\le\frac{9}{100}\qquad \Longrightarrow\qquad \e_{gap}\le \begin{cases} 
1\,,\ \lgb>0\,,\\
\frac{16}{27}\,,\ \lgb<0\,.
\end{cases}
\end{equation}
Additionally, it was pointed out   \cite{Camanho:2014apa}
that pure EGB gravity with a negative cosmological constant can not arise as a low-energy limit of a gauge theory/string theory 
correspondence ---
the difference of central charges $(c-a)/c$ is bounded by $\Delta_{gap}^{-2}$, where $\Delta_{gap}$ is the dimension of the lightest 
single particle operators with spin $J>2$ in the holographically dual conformal gauge theory. 
Integrating out massive 
$J>2$ spin states generically produces new higher-curvature contributions, in addition to the Gauss-Bonnet term.
These higher curvature corrections are as important as the Einstein-Hilbert term and the GB term in 
\eqref{eq:aisnotc} when the size of a black hole becomes of order $\lgb L$. The latter is 
true even as $\lgb \ll 1$, as the Ricci scalar evaluated on the horizon of $\sim \lgb L$ size black hole
\eqref{eq:statbh} diverges as $\frac 1\lgb$.

\section{PW/BEFP effective actions}\label{action}

We begin with description of the PW effective action \cite{pw}. 
The action of the effective five-dimensional supergravity including the
scalars $\alpha$ and $\chi$ (dual to mass terms for the bosonic and
fermionic components of the hypermultiplet respectively) is given by
\begin{equation}
\begin{split}
S=&\,
\int_{\calm_5} d\xi^5 \sqrt{-g}\ \call_{PW}\\
=&\frac{1}{4\pi G_5}\,
\int_{\calm_5} d\xi^5 \sqrt{-g}\left[\ft14 R-3 (\del\a)^2-(\del\chi)^2-
\calp\right]\,,
\end{split}
\eqlabel{action5}
\end{equation}
where the potential%
\footnote{We set the five-dimensional 
supergravity coupling to one. This corresponds to setting the
radius $L$ of the five-dimensional sphere in the undeformed metric
to $2$.}
\begin{equation}
\calp=\frac{1}{16}\left[\frac 13 \left(\frac{\del W}{\del
\a}\right)^2+ \left(\frac{\del W}{\del \chi}\right)^2\right]-\frac
13 W^2\,,
 \eqlabel{pp}
\end{equation}
is a function of $\alpha$ and $\chi$, and is determined by the
superpotential
\begin{equation}
W=- e^{-2\alpha} - \frac{1}{2} e^{4\alpha} \cosh(2\chi)\,.
\eqlabel{supp}
\end{equation}
In our conventions, the five-dimensional Newton's constant is
\begin{equation}
G_5\equiv \frac{G_{10}}{2^5\ {\rm vol}_{S^5}}=\frac{4\pi}{N^2}\,.
\eqlabel{g5}
\end{equation}
Supersymmetric vacuum of $\caln=2^*$ gauge theory in Minkowski space-time is given 
by 
\begin{equation}
ds_5^2=e^{2 A}\left(-dt^2 +d\vec{x}^2\right)+dr^2\,,\qquad \r=\r(r)\equiv e^{\a(\r)}\,,\qquad \chi=\chi(r)\,,
\eqlabel{pwsusy}
\end{equation}
with 
\begin{equation}
\begin{split}
e^A&=\frac{k \r^2}{\sinh(2\chi)}\,,\qquad \r^6=\cosh(2\chi)+\sinh^2(2\chi)\,\ln\frac{\sinh(\chi)}{\cosh(\chi)}\,,
\qquad \frac{dA}{dr}=-\frac 13 W\,,
\end{split}
\eqlabel{pwsolution}
\end{equation}
where the single integration constant $k$ is related to the hypermultiplet
mass $m$ according to \cite{bpp}
\begin{equation}
k= m L =2 m\,.
\eqlabel{kim}
\end{equation}

The BEFP effective action \cite{Bobev:2013cja} is 
given by 
\begin{equation}
\begin{split}
S_{BEFP}=&\,
\int_{\calm_5} d\xi^5 \sqrt{-g}\ \call_{BEFP}\\
=&\frac{1}{4\pi G_5}\,
\int_{\calm_5} d\xi^5 \sqrt{-g}\left[ R-12 \frac{(\del\eta)^2}{\eta^2}
-4 \frac{(\del{\vec X})^2}{(1-\vec{X}^2)^2}
-\calv\right]\,,
\end{split}
\eqlabel{befp}
\end{equation}
with the potential 
\begin{equation}
\calv=-\left[\frac{1}{\eta^4}+2\eta^2\ \frac{1+\vec{X}^2}{1-\vec{X}^2}
-\eta^8\ \frac{(X_1)^2+(X_2)^2}{(1-\vec{X}^2)^2}
\right]\,,
 \eqlabel{pbefp}
\end{equation}
where $\vec{X}=\left(X_1,X_2,X_3,X_4,X_5\right)$ are five of the scalars and $\eta$ is the sixth. 
The symmetry of the action reflects the symmetries of the dual gauge theory: the 
two scalars $(X_1,X_2)$ form a doublet under the $U(1)_R$ part of the gauge group, 
while $(X_3,X_4,X_5)$ form a triplet under $SU(2)_V$ and $\eta$ is neutral.  
The PW effective action is recovered as a consistent truncation of \eqref{befp} with 
\begin{equation}
X_2=X_3=X_4=X_5=0\,,
\eqlabel{truncate}
\end{equation}
provided we identify the remaining BEFP scalars $(\eta,X_1)$ with the PW scalars $(\a,\chi)$ as follows
\begin{equation}
e^\a\equiv \eta\,,\qquad \cosh 2\chi =\frac{1+(X_1)^2}{1-(X_1)^2} \,.
\eqlabel{id}
\end{equation}
Note that once $m\ne 0$ (correspondingly $X_1\ne 0$), the $U(1)_R$ symmetry is
explicitly broken; on the contrary, $SU(2)_V$ remains unbroken in truncation to PW.

\section{Holographic duals to $\caln=2^*$ vacuum states on $S^3$}\label{vacuum}

We derive bulk equations of motion and specify boundary conditions representing gravitational 
dual to vacuum states of  strongly coupled $\caln=2^*$ gauge theory on $S^3$. We assume that the vacua are  
$SO(4)$-invariant. We argue that there is no obstruction of exciting these vacua by arbitrarily small perturbations 
of the bulk scalar fields $\a$ and $\c$.  
We review holographic renormalization of the 
theory and compute the vacuum energy. Next, we solve  static gravitational equations 
perturbatively in the mass deformation parameter $m\ell\ll 1$ --- this would serve 
as an independent check for the general $\calo(m\ell)$  numerical solutions.  
We conclude with the plot representing $\e\equiv E_{vacuum}(m\ell)/E_{vacuum}^{\caln=4}$,
\begin{equation}
E_{vacuum}^{\caln=4}\equiv E_{vacuum}(m\ell=0)= \frac{3N^2}{16\ell}\,,
\eqlabel{en4}
\end{equation}
as  a function of $m\ell$. Interestingly, while the vacuum energy of the $\caln=4$ SYM is positive, 
it is negative\footnote{Prior to imposing causality 
constraints in EGB gravity, its vacuum energy becomes negative once $\lgb> 5/36$. Vacuum energy of a different nonconformal 
gauge theory on $S^3$ was also observed to be negative 
in  \cite{Buchel:2011cc}.} 
for $\caln=2^*$ gauge theory once $m\ell\gtrsim 0.87$.

\subsection{Equations of motion and the boundary conditions}
We consider the general time-dependent $SO(4)$-invariant 
ansatz for the metric and the scalar fields:
\begin{equation}
ds_5^2=\frac{4}{\cos^2 x} \left(-A e^{-2\dd} (dt)^2+\frac{(dx)^2}{A}+\sin^2 x (d\Omega_3)^2\right)\,,
\eqlabel{geomdyn}
\end{equation}
where $(d\Omega_3)^2$ is a metric on a unit\footnote{We set $\ell=1$; the $\ell$ 
dependence can be easily recovered from dimensional analysis.} round $S^3$, and $\{A,\dd,\a,\chi\}$ being functions of a radial coordinate 
$x$ and time $t$. Introducing 
\begin{equation}
\Phi_\a\equiv \del_x \a\,,\ \ \Phi_\c\equiv \del_x \c\,,\ \ \Pi_\a\equiv \frac{e^\dd}{A}\del_t\a\,,\ \  \Pi_\c\equiv \frac{e^\dd}{A}\del_t\c\,,
\eqlabel{momenta}
\end{equation}
we obtain from \eqref{action5} the following equations of motion:
\nxt the evolution equations, $ \dot{} = \del_t$,
\begin{equation}
\begin{split}
&\dot{\a}=A e^{-\dd} \Pi_\a\,,\qquad \dot{\c}=A e^{-\dd} \Pi_\c\,,\\
&\dot\Phi_\a=\left(A e^{-\dd} \Pi_\a\right)_{, x}\,,\qquad \dot\Phi_\c=\left(A e^{-\dd} \Pi_\c\right)_{, x}\,,\\
&\dot\Pi_\a=\frac{1}{\tan^3 x}\left(\tan^3x A e^{-\dd}\Phi_\a\right)_{,x}-\frac {2 }{3\cos^2 x}e^{-\dd} \frac{\del\calp}{\del\a}\,,\\
&\dot\Pi_\c=\frac{1}{\tan^3 x}\left(\tan^3x A e^{-\dd}\Phi_\c\right)_{,x}-\frac {2 }{\cos^2 x}e^{-\dd} \frac{\del\calp}{\del\c}\,,  
\end{split}
\eqlabel{evolution}
\end{equation}
\nxt the spatial constraint equations,
\begin{equation}
\begin{split}
A_{,x}=&\frac{2+2\sin^2 x}{\sin x\cos x}(1-A)-2\sin (2x) A \left(\Phi_\a^2+\Pi_\a^2+\frac 13 \Phi_\c^2
+\frac 13 \Pi_\c^2\right)\\&-4\tan x \left(1+\frac 43 \calp\right)\,,\\
\dd_{, x}=&-2\sin(2x) \left(\Phi_\a^2+\Pi_\a^2+\frac 13 \Phi_\c^2
+\frac 13 \Pi_\c^2\right)\,,
\end{split}
\eqlabel{sconstrant}
\end{equation}
\nxt and the moment constraint equation, 
\begin{equation}
\begin{split}
A_{,t}+4 \sin(2x) A^2 e^{-\dd} \left(\Phi_\a\Pi_\a+\frac 13 \Phi_\c\Pi_\c\right)=0\,.
\end{split}
\eqlabel{momconstraint}
\end{equation}
It is straightforward to verify that the spatial derivative of \eqref{momconstraint} 
is implied by \eqref{evolution} and \eqref{sconstrant}; thus is it sufficient 
to impose this equation at a single point. As $x\to 0_+$, the momentum 
constraint implies that $A(0,t)$ is a constant\footnote{In fact, 
the non-singularity of $A(t,x)$ in this limit automatically solves \eqref{momconstraint}.}, and as $x\to \frac{\pi}{2}_-$    
the latter constraint is equivalent to the conservation of the boundary stress-energy tensor
(see \ref{holren} for details). 

The general non-singular solution of \eqref{evolution}, \eqref{sconstrant} at the origin takes form 
\begin{equation}
\begin{split}
&A(t,x)=1+\calo(x^2)\,,\qquad \dd(t,x)=d^h_0(t)+\calo(x^2)\,,\\
&\a(t,x)=\a^h_0(t)+\calo(x^2)\,,\qquad  \c(t,x)=\c^h_0(t)+\calo(x^2)\,.
\end{split}
\eqlabel{dynir}
\end{equation}
It is completely characterized by three time-dependent functions: 
\begin{equation}
\{d^h_0\,, \a^h_0\,, \c^h_0\}\,.
\eqlabel{ircoefficients}
\end{equation}
At the outer boundary $x=\frac \pi2$ we introduce $y\equiv \cos^2 x$ so that we have
\begin{equation}
\begin{split}
A=&1+y\ \frac 23  c_{1,0}\  + y^2\ \biggl(a_{2,0}(t)+\biggl(\frac 23 c_{1,0}(c_{1,0}+1)+8\r_{1,1}^2+
16\r_{1,1}\r_{1,0}(t)\biggr)\ln y\\
&+8 \r_{1,1}^2\ln^2 y\biggr)+ \calo(y^3\ln^3 y)\,,\\
\dd=&y\ \frac13  c_{1,0}+y^2\ \biggl(\frac12 c_{2,0}(t)-\frac{1}{36} c_{1,0}^2+4 \r_{1,0}^2(t)-\frac18 c_{1,0}+2 \r_{1,1}^2
+4 \r_{1,0}(t) \r_{1,1}\\
&+\biggl(\frac 14 c_{1,0}+\frac13 c_{1,0}^2+4 \r_{1,1}^2+8 \r_{1,0}(t) \r_{1,1}\biggr) \ln y
+4 \r_{1,1}^2 \ln^2 y\biggr)+ \calo(y^3\ln^3 y)\,,\\
e^\a=&1+y\ \left(\r_{1,0}(t)+\r_{1,1} \ln y\right)+y^2\ 
\biggl(
\frac{1}{12} c_{1,0}^2+\r_{1,0}(t)-3 \r_{1,1} c_{1,0}+6 \r_{1,1}^2\\
&-4 \r_{1,0}(t) \r_{1,1}+\frac43 c_{1,0} \r_{1,0}(t)+\frac32 \r_{1,0}^2(t)
+\frac14 \del^2_{tt}\r_{1,0}(t)
+\biggl(\frac43 \r_{1,1} c_{1,0}+\r_{1,1}\\&-4 \r_{1,1}^2
+3 \r_{1,0}(t) \r_{1,1}\biggr) \ln y+\frac32 \r_{1,1}^2 \ln^2 y
\biggr)+\calo(y^3\ln^3 y)\,,\\
\cosh2\c=&1+y\ c_{1,0}+y^2\ \biggl(c_{2,0}(t)+\biggl(\frac12 c_{1,0}+\frac23 c_{1,0}^2\biggr) \ln y\biggr)+ \calo(y^3\ln^2 y)\,,
\end{split}
\eqlabel{dynuv}
\end{equation}
where we explicitly indicated time-dependence, \ie 
\begin{equation}
\frac{d}{dt} c_{1,0}=0\,,\qquad \frac{d}{dt} \r_{1,1}=0\,.
\eqlabel{timedep}
\end{equation}
Asymptotic expansion \eqref{dynuv} is completely characterized by two 
constants\footnote{Prescribing  time dependence to these coefficients 
amounts to study quantum quenches in $\caln=2^*$ gauge theory \cite{Buchel:2012gw}.} 
$\{\r_{1,1},c_{1,0}\}$ and three time-dependent functions 
\begin{equation}
\{a_{2,0}\,, \r_{1,0}\,, c_{2,0}\}\,,
\eqlabel{uvcoefficients}
\end{equation}
constraint by  \eqref{momconstraint} to satisfy 
\begin{equation}
0=\frac{d}{dt}\biggl(a_{2,0}-8\r_{1,0}^2(t)-16\r_{1,0}(t)\r_{1,1}-\frac 23 c_{2,0}(t)\biggr)\,.
\eqlabel{uvmom}
\end{equation}
The non-normalizable coefficients $\r_{1,1}$ and $c_{1,0}$ 
are related to the mass deformation parameters of the dual gauge theory. 
Following \cite{Buchel:2007vy}, the precise relation can be established by matching the 
asymptotics \eqref{dynuv} with the supersymmetric PW RG flow  \eqref{pwsolution},
\begin{equation}
\{\r_{1,1},c_{1,0}\}\bigg|_{PW}=k^2\ \left\{\frac {1}{48},\frac 18\right\}=m^2\ \left\{\frac {1}{12},\frac 12\right\}\,.
\eqlabel{matchPW}
\end{equation}
A specific relation between the non-normalizable coefficients of the bulk scalars $e^\a$ and $\cosh2\c$, \ie
\begin{equation}
c_{1,0}=6\r_{1,1}\,,
\eqlabel{n2susy}
\end{equation} 
realizes $\caln=2$ supersymmetry of the boundary gauge theory in the UV. 
As in \cite{Buchel:2007vy}, it is possible to study the theory with explicitly broken supersymmetry, \ie
\begin{equation}
\r_{1,1}\equiv\frac{1}{48}\ (m_b L)^2\qquad \ne\qquad  \frac 16\ \times\ c_{1,0}\equiv \frac 16\ \times\ \frac 18\ (m_fL)^2  \,,
\eqlabel{genflow}
\end{equation}  
where $m_b$ and $m_f$ are the masses of the bosonic and the fermionic components of the 
$\caln=2$ hypermultiplet of the boundary gauge theory.  

A non-equilibrium state of the gauge theory can be specified with the following initial/boundary conditions:
\begin{equation}
\begin{split}
&\a(0,x)=\a^{init}(x)\,,\qquad  \c(0,x)=\c^{init}(x)\,,\qquad 
\Phi_\a(0,x)=\Phi_\a^{init}=\frac{d\a^{init}}{dx}\,,\\
& \Phi_\c(0,x)=\Phi_\c^{init}=\frac{d\c^{init}}{dx}\,,\qquad \Pi_\a(0,x)=\Pi_\a^{init}(x)\,,\qquad \Pi_\c(0,x)=\Pi_\c^{init}(x)\,,
\end{split}
\eqlabel{init1}
\end{equation}
and as  $y\equiv \cos^2 x\to 0$,
\begin{equation}
\begin{split}
&\a^{init}(y)= \r_{1,1}\ y\ln y +\calo(y)\,,\qquad \cosh\left(2 \c^{init}(y)\right)
=1+ y\ c_{1,0}+\calo(y^2 \ln y)\,,\\
&\Pi_\a^{init}(y)=\calo(y)\,,\qquad \Pi_\c^{init}(y)=\calo(y^{3/2})\,,
\end{split}
\eqlabel{init2}
\end{equation}
\begin{equation}
\begin{split}
A(0,x)=&1+\frac{\cos^4 x}{\sin^2 x}\ \exp\biggl(-\frac 23 \int_0^x d\xi\ \sin(2\xi)\biggl(
\left(\Pi_c^{init}(\xi)\right)^2+\left(\Phi_c^{init}(\xi)\right)^2\\
&+3
\left(\Pi_\a^{init}(\xi)\right)^2+3 \left(\Phi_\a^{init}(\xi)\right)^2
\biggr)
\biggr)\ \times\ g(x)\,,\\
g(x)=&-\frac43\int_0^x d\xi \tan^3\xi \exp\biggl(\frac 23 \int_0^\xi d\eta\ \sin(2\eta)\biggl(
\left(\Pi_c^{init}(\eta)\right)^2+\left(\Phi_c^{init}(\eta)\right)^2\\
&+3
\left(\Pi_\a^{init}(\eta)\right)^2+3 \left(\Phi_\a^{init}(\eta)\right)^2
\biggr)
\biggr)\ \times\ \biggl(\frac{4\calp^{init}(\xi)+3}{\cos^2\xi}+\left(\Pi_c^{init}(\xi)\right)^2\\
&+\left(\Phi_c^{init}(\xi)\right)^2
+3\left(\Pi_\a^{init}(\xi)\right)^2+3 \left(\Phi_\a^{init}(\xi)\right)^2\biggr)\\
&\calp^{init}(\xi)=\calp\left(\a^{init}(\xi),\c^{init}(\xi)\right)\,,
\end{split}
\eqlabel{init3}
\end{equation}
\begin{equation}
\begin{split}
&\dd(0,x)=-\frac 23 \int_0^x d\xi\ \sin(2\xi)\biggl(
\left(\Pi_c^{init}(\xi)\right)^2+\left(\Phi_c^{init}(\xi)\right)^2
+3
\left(\Pi_\a^{init}(\xi)\right)^2+3 \left(\Phi_\a^{init}(\xi)\right)^2
\biggr)\,,
\end{split}
\eqlabel{init4}
\end{equation}
where we explicitly solved for $A(0,x)$ and $\dd(0,x)$ using constraint equations \eqref{sconstrant}.
Notice that while $A(0,x)$ and $\dd(0,x)$ are free from the singularities given arbitrary profiles 
\eqref{init1}, a large amplitude initial conditions might cause $A(0,x)$ to vanish 
for some  $0<x_0<\frac {\pi}{2}$, \ie $A(0,x_0)=0$, --- this corresponds to 'putting a black hole in the initial data'. 
Clearly, initial conditions arbitrarily small perturbed about static gravitational solutions without a horizon (see below) 
are well defined. In particular one can can consider perturbations with  
\begin{equation}
\a^{init}=\a^{v}\,,\quad \c^{init}=\c^{v}\,,\quad \Pi_{\a,\c}^{init} =\l\ \pi_{\a,\c}(x)\,,\qquad \l\to 0 \,,
\eqlabel{momonly}
\end{equation}
where the superscript $ ^{v}$ stands for a  static (vacuum) solution and $\l$ characterizes an overall amplitude 
of the perturbation with  given initial profiles $\pi_\a$ and $\pi_\c$.

The $SO(4)$-invariant  vacua of strongly coupled $\caln=2^*$ gauge theory correspond to static 
solutions of \eqref{evolution}-\eqref{momconstraint}. To avoid unnecessary cluttering of the formulas, 
we omit the superscript $ ^v$,  use a radial coordinate $y\equiv \cos^2 x$, and introduce  
\begin{equation}
A(t,y)=a(y)\,,\qquad \dd(t,y)=d(y)\,,\qquad e^{\a(t,y)}=\r(y)\,,\qquad \cosh(2\c(t,y))=c(y)  \,.
\eqlabel{static}
\end{equation}
We find then 
\begin{equation}
\begin{split}
&0=c''-\frac{c (c')^2}{c^2-1}+c' \left(\frac{a'}{a}-d'\right)
-\frac{(y+1) c'}{y (1-y)}- \frac{\r^2 (c^2-1) (\r^6 c-4)}{4(1-y) y^2 a}\,,\\
&0=\r''-\frac{(\r')^2}{\r}+\r' \left(\frac{a'}{a}-d'\right)-\frac{(y+1) \r'}{y (1-y)}
-\frac{(c^2-1) \r^9}{12(1-y) y^2 a}-\frac{1-\r^6 c}{6\r^3 y^2 a (1-y)}\,,\\
&0=d'-\frac{2 y (1-y) (c')^2}{3(c^2-1)}-\frac{8 (1-y) y (\r')^2}{\r^2}\,,\\
0&=a'-(y-y^2)a\left(\frac{8    (\r')^2}{\r^2}+\frac{2  (c')^2}{3(c^2-1)}\right)+\frac{(y-2) a+y}{y (1-y)}
-\frac{(c^2-1) \r^8-8\r^2 c}{6y}+\frac{2}{3 y \r^4}\,,
\end{split}
\eqlabel{vaceoms}
\end{equation}
where $ '=\frac{d}{dy}$. The boundary conditions as $y\to 0$ are as in \eqref{dynuv}, once we neglect the time dependence. 
At the origin, using $z\equiv 1-y$ we have
\begin{equation}
\begin{split}
&a=1+\left(-1+\frac{1}{3(\r^h_0)^4}-\frac{(\r^h_0)^8}{12}\left((c^h_0)^2-1\right)+\frac{2c^h_0(\r^h_0)^2}{3}\right)\ z+\calo(z^2)\,,\\
&d=d^h_0+\calo(z^2)\,,\\
&\r=\r^h_0+\left(\frac{(\r^h_0)^9}{24}\left((c^h_0)^2-1\right)+\frac{1-(\r^h_0)^6c^h_0}{12 (\r^h_0)^3}\right)\ z+\calo(z^2)\,,\\
&c=c^h_0+\frac18(\r^h_0)^2\left((c^h_0)^2-1\right)\left(c^h_0(\r^h_0)^6-4\right)\ z+\calo(z^2)\,.
\end{split}
\eqlabel{originz}
\end{equation}
We consider geometries with $\caln=2$ supersymmetry in the ultraviolet, so we impose the constraint \eqref{matchPW}.
Having fixed $m$, the complete set of normalizable coefficients in the UV/IR is given by:
\begin{equation}
\{a_{2,0}\,,\ \r_{1,0}\,,\ c_{2,0}\,,\ \r^h_0\,,\ c^h_0\,,\ d^h_0\}\,.
\eqlabel{fullset}
\end{equation} 
Note that the six integration constants \eqref{fullset} is exactly what is needed to uniquely fix a solution
of a coupled system of two second-order  and  two  first-order ODEs.

\subsection{Holographic renormalization and the vacuum energy}\label{holren}
Holographic renormalization of RG flows in PW geometry was discussed in 
\cite{Buchel:2004hw}. Here we apply the analysis for the gravitational solutions dual 
to vacua of $\caln=2^*$ gauge theory on $S^3$.  

The gravitational action \eqref{action5} evaluated 
on a static solution \eqref{vaceoms} diverges 
--- this divergence is a gravitational reflection of 
a standard UV divergence of the free energy in the interacting 
boundary gauge theory. It is regulated by cutting off the radial 
coordinate integration at $y=y_c\ll 1$.   
It is straightforward to verify that the regularized Euclidean  
gravitational Lagrangian, $\call_{reg}^E$, is a total derivative, 
\begin{equation}
\begin{split}
\call_{reg}^E=&\frac{1}{4\pi G_5} {\rm vol(\Om_3)} 
\int_{1}^{y_c} dy\ \frac{d}{dy}
\left(\frac{4(1-y)^2e^{-d}}{y^2}\ \left(a+2y a d'-y a' 
\right)\right)\\
=&\frac{\rm{vol} (\Om_3)}{4\pi G_5}\  
\left[\frac{4(1-y)^2e^{-d}}{y^2}\ \left(a+2y a d'-y a' 
\right)\right]\bigg|^{y_c}\,,
\end{split}
\eqlabel{totder}
\end{equation}
where in the second equality, using \eqref{originz}, we observe that 
the only contribution comes from the upper limit of integration.
Regularized Lagrangian \eqref{totder} has to be supplemented with contributions 
coming from the familiar Gibbons-Hawking term, $\call_{GH}^E$, 
\begin{equation}
\begin{split}
S_{GH}^E=&-\frac{1}{8\pi G_5}\int_{\del\calm_5}d\xi^4 \sqrt{h_E}\nabla_\mu n^\mu
\equiv \int dt_E \call_{GH}^E\,,\\
\call_{GH}^E=&\frac{\rm{vol}(\Om_3)}{4\pi G_5}\biggl[
\frac{4 (1-y) e^{-d}}{y^2} \left(a (y-4) -2 d' y(1-y) a
+a' y(1-y)\right)
\biggr]\bigg|^{y_c}\,,
\end{split}
\eqlabel{lgh} 
\end{equation}
and the counterterm Lagrangian\footnote{We keep only the 
counterterms relevant for the $R\times S^3$ background geometry 
of the gauge theory.}, $\call_{counter}^E$, 
\begin{equation}
\begin{split}
S_{counter}^E\equiv& \int dt_E \call_{counter}^E\,,\\
\call_{counter}^E=&\frac{\rm{vol}\Om_3}{4\pi G_5}\sqrt{h_E} \biggl[
\frac 34+\frac 14 R_4+\frac 12 \c^2 +3\a^2-\frac 32 \frac{\a^2}{\ln\e_c}
\\&+\ln\e_c \left(-\frac 13 \c^2 R_4-\frac 23 \c^4\right)+\frac 16\c^4
\biggr]\bigg|^{y_c}\,,
\end{split}
\eqlabel{counter}
\end{equation}
where $R_4\equiv R_4(h_E)$ is the Ricci scalar constructed from $h_E$,
and $\e_c$ parameterizes conformal anomaly terms in terms of the $g_{t_Et_E}$
metric component, 
\begin{equation}
R_4=\frac{3y}{2(1-y)}\,,\qquad \e_c\equiv \sqrt{g_{t_Et_E}}
=\frac{2\sqrt{a}e^{-d}}{\sqrt{y}}\,. 
\eqlabel{r4}
\end{equation}  
The renormalized Lagrangian $\call_{renom}^E$, finite in the limit 
$y_c\to 0$, is identified with the free energy $\calf$ 
of the boundary gauge theory,   
\begin{equation}
\begin{split}
&\calf=\call_{renom}^E=\lim_{y_c\to 0}\biggl(
\call^E_{reg}+\call^E_{GH}+\call^E_{counter} \biggr)\\
=&\frac{\rm{vol}\Om_3}{4\pi G_5}\ \frac 32\ 
\biggl(1+c_{1,0}^2 \left(\frac 49-\frac{16}{9}\ln 2\right)
+c_{1,0}\left(-\frac 43-\frac83 \ln 2\right)+ 64 \r_{1,1}^2\ln 2
\\&+\biggl\{64 \r_{1,1}\r_{1,0}+\frac 83 c_{2,0}+32 \r_{1,0}^2-4 a_{2,0}\biggr\}
\biggr)\\
=&\frac{3N^2}{16\ell}\biggl(1+\frac{(m\ell)^4}{9}-\frac 23 (1+2\ln 2)(m\ell)^2
+\biggl\{
32 \r_{1,0}^2+\frac{16}{3}(m\ell)^2 \r_{1,0}+\frac 83 c_{2,0}-4a_{2,0}
\biggr\}
\biggr)\,,
\end{split}
\eqlabel{freeenergy}
\end{equation} 
where in the second line we used the asymptotic expansion \eqref{dynuv} 
and expressed the last line in terms of gauge theory variables 
using \eqref{g5} and  \eqref{matchPW} and restoring the size $\ell$
of the $S^3$. Several comments are in order:
\nxt For static gravitational solutions without Schwarzschild horizon
(as discussed here), the free energy $\calf$ 
must coincide with the  energy $E$ of the boundary stress-energy tensor. 
We explicitly verified that, indeed, 
\begin{equation}
\calf=E\equiv E_{vacuum}(m\ell)\,.
\end{equation}  
The latter is identified with the vacuum energy of $\caln=2^*$ gauge theory on $S^3$.
\nxt In a limit when all the (non-)normalizable coefficients vanish we recover the 
vacuum energy of the $\caln=4$ SYM \eqref{en4}.
\nxt It is easy to extend discussion for general $SO(4)$-invariant 
non-equilibrium states of $\caln=2^*$ gauge theory --- the final answer is as 
\eqref{freeenergy}, except with $\{\r_{1,0}\,, c_{2,0}\,, a_{2,0}\}$ now  being functions of time. 
Note that 
\begin{equation}
\frac{d\cale}{dt}\propto  \frac{d}{dt}\ \biggl(\ 4 \biggl\{16 \r_{1,1}\r_{1,0}(t)+\frac 23 c_{2,0}(t)+8 \r_{1,0}^2(t)- a_{2,0}(t)\biggr\}\ \biggr)=0\,,
\eqlabel{conserve}
\end{equation}
according to \eqref{uvmom}.
That is, the boundary gauge theory energy conservation is enforced by the bulk  momentum constraint \eqref{momconstraint}.

\subsection{Vacuum states for $m\ell\ll 1$}\label{vsmall}
In preparation to the full numerical solution of \eqref{vaceoms}, we discuss here its perturbative solution for $\r_{1,1}\ll 1$. 
We introduce 
\begin{equation}
\begin{split}
&c=\cosh(2\l \chi_1(y)+\calo(\l^3))\,,\qquad \r=e^{\l^2 \a_2(y)+\calo(\l^4)}\,,\\
&a=1+\l^2 a_2(y)+\calo(\l^4)\,,\qquad d=\l^2 d_2(y)+\calo(\l^2)\,,
\end{split}
\eqlabel{pertvacuum}
\end{equation}
where $\l$ is a small parameter.  Substituting \eqref{pertvacuum} into \eqref{vaceoms} we find
\begin{equation}
\begin{split}
&0=\c_1''-\frac{1+y}{y(1-y)} \c_1' +\frac {3}{4y^2(1-y)}\c_1\,,\\
&0=\a_2''-\frac{1+y}{y(1-y)} \a_2' +\frac {1}{y^2(1-y)}\a_2\,,\\
&0=a_2'-\frac{2-y}{y(1-y)}a_2-\frac 83 y(1-y)(\c_1')^2+\frac 2y(\c_1)^2\,,\\
&0=d_2'-\frac 83 y(1-y)(\c_1')^2\,.
\end{split}
\eqlabel{perteoms}
\end{equation}
Solutions to \eqref{perteoms} must satisfy boundary conditions corresponding to \eqref{dynuv} and
\eqref{originz}. We can solve equation for $\a_2$ analytically, 
\begin{equation}
\a_2=\r_{1,1,(2)}\ \frac{y\ln y }{1-y}\,,
\eqlabel{al2anal}
\end{equation}
where $\r_{1,1,(2)}$ is the non-normalizable integration coefficient. The remaining equations in \eqref{perteoms}
are solved with ``shooting method'' developed in \cite{Aharony:2007vg}.  In particular, given 
the asymptotic expansions in the UV, $y\to 0_+$,  
\begin{equation}
\begin{split}
\c_1=&y^{1/2} \left(1+y\ \left(\c_{1,0,(1)} +\frac 14 \ln y\right)+\calo(y^2\ln y)\right)\,,\\
a_2=&\frac 43 y + y^2\ \left( a_{2,0,(2)}+\frac 43\ln y\right)+\calo(y^3\ln^2 y)\,,\\
d_2=&\frac 23 y +y^2\ \left(-\frac 14 +2 \c_{1,0,(1)} +\frac 12 \ln y\right)+\calo(y^3\ln^2 y)\,,
\end{split}
\eqlabel{pertuv}
\end{equation}
and in the IR, $z\to 0_+$,
\begin{equation}
\begin{split}
\c_1=&\c_{0,(1)}^{h}\left(1-\frac 38 z+\calo(z^2)\right)\,,\\
a_2=&(\c_{0,(1)}^{h})^2\ \left(z-\frac 58 z^2 +\calo(z^3)\right)\,,\\
d_2=&d^h_{0,(2)}-\frac{3}{16}(\c_{0,(1)}^{h})^2 z^2 +\calo(z^3)\,,
\end{split}
\eqlabel{pertir}
\end{equation}
we find  numerically,
\begin{equation} 
\begin{tabular}{ c | c |c| c }
 $\c_{1,0,(1)}$ &  $a_{2,0,(2)}$ & $\c_{0,(1)}^{h}$ &  $d^h_{0,(2)}$\\
\hline
  0.0568528 & -0.363452  & 0.785398 & 0.199266
\end{tabular}\,.
\eqlabel{table1}
\end{equation} 
To compare with the full numerical solution, we identify, to order $\calo(\l^2)$, 
\begin{equation}
\begin{split}
&\r_{1,1}=\r_{1,1,(2)} \l^2\,,\qquad c_{1,0}=2 \l^2\,,\qquad \r_{1,0}=0\,,\qquad c_{2,0}=4\c_{1,0,(1)} \l^2\,, \\
&a_{2,0}=a_{2,0,(2)}\l^2\,,\qquad \r^h_0=1-\r_{1,1,(2)}\l^2\,,\qquad c^h_0=1+2(\c_{0,(1)}^{h})^2\l^2\,,\qquad 
d^h_0=d^h_{0,(2)}\l^2\,.
\end{split}
\eqlabel{pertid}
\end{equation}
Note that $\caln=2$ supersymmetry in the UV at $\calo(\l^2)$ leads to (see \eqref{n2susy})
\begin{equation}
\r_{1,1,(2)}=\frac 13 \,.
\eqlabel{r112}
\end{equation}
From \eqref{freeenergy},
\begin{equation}
\begin{split}
\e\equiv \frac{E_{vacuum}}{E_{vacuum}^{\caln=4}}=&
1+\left(\frac{32}{3}\c_{1,0,(1)}-4a_{2,0,(2)}-\frac 83(1+2\ln 2)\right)\l^2+\calo(\l^4)\\
=&1+\left(\frac{8}{3}\c_{1,0,(1)}-a_{2,0,(2)}-\frac 23(1+2\ln 2)\right)(m\ell)^2+\calo((m\ell)^4)\,.
\end{split}
\eqlabel{epvacuum}
\end{equation}

\subsection{Gravitational solution and $E_{vacuum}$ for general $m\ell$}

\begin{figure}[t]
\begin{center}
\psfrag{x}{{$\r_{1,1}$}}
\psfrag{y1}{{$c_{2,0}$}}
\psfrag{y2}{{$\r_{1,0}$}}
\psfrag{y3}{{$a_{2,0}$}}
\psfrag{y4}{{$\r_{0}^h$}}
\psfrag{y5}{{$c_{0}^h$}}
\psfrag{y6}{{$d_{0}^h$}}
\includegraphics[width=2.6in]{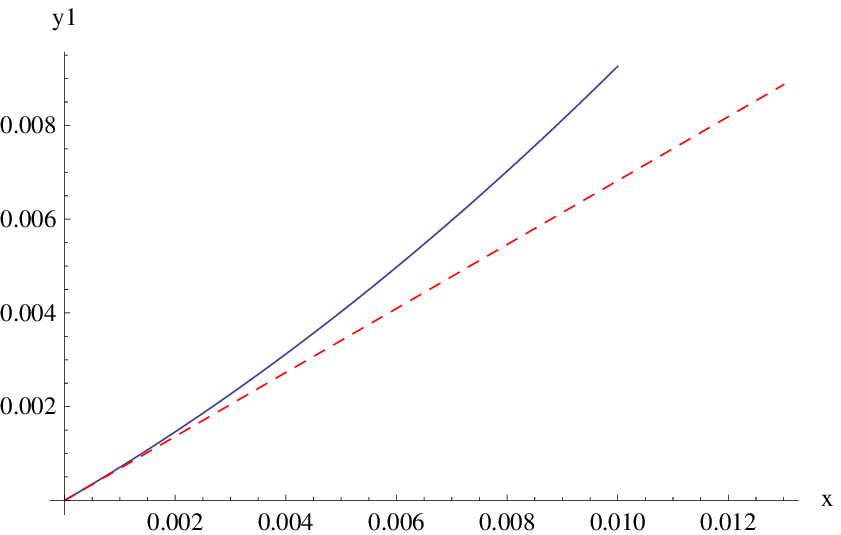}\qquad
\includegraphics[width=2.6in]{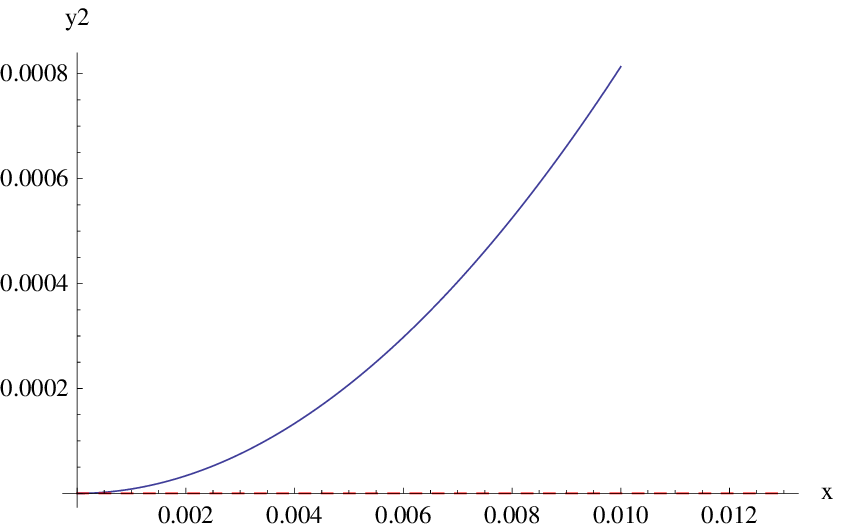}
\vskip 0.2cm
\includegraphics[width=2.6in]{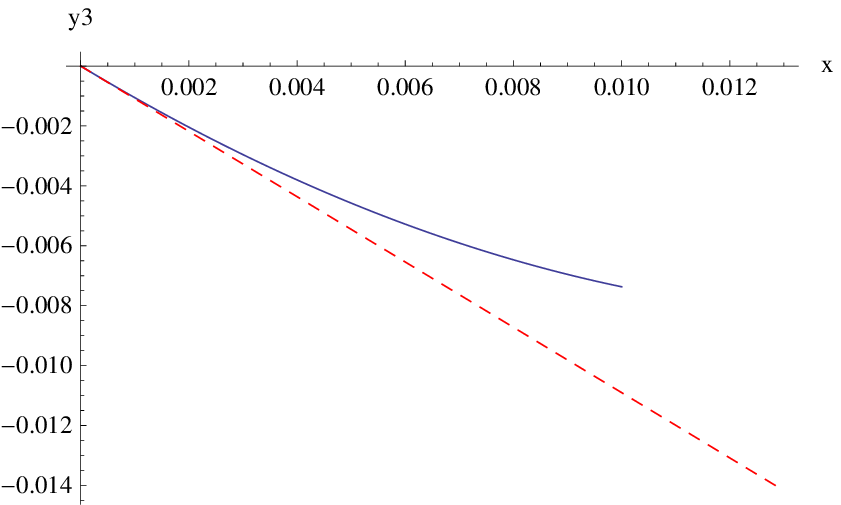}\qquad
\includegraphics[width=2.6in]{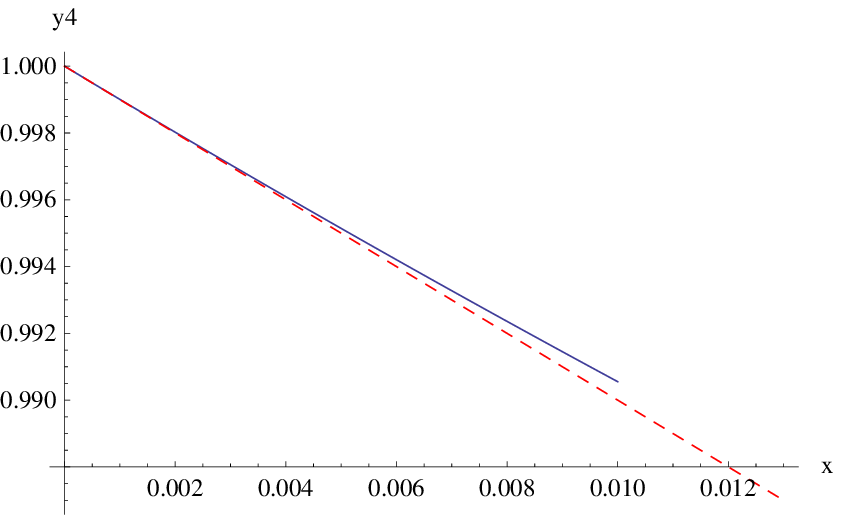}\qquad
\vskip 0.2cm
\includegraphics[width=2.6in]{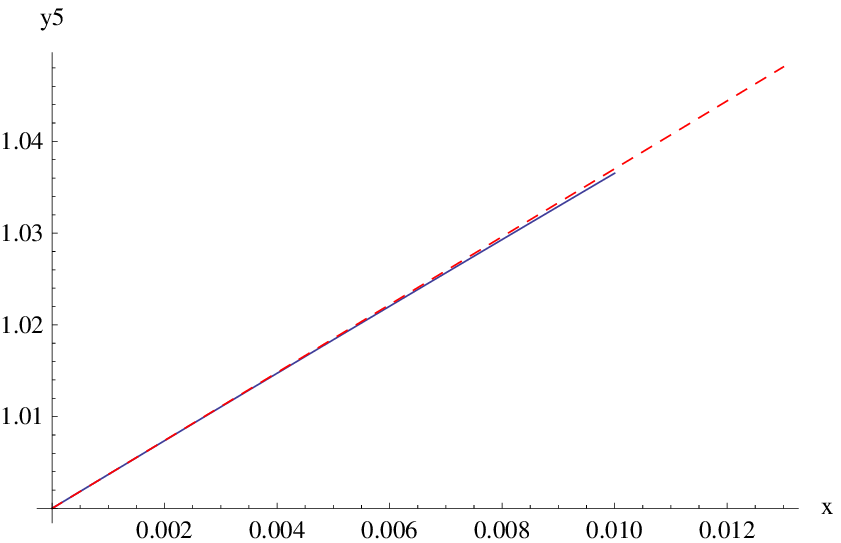}\qquad
\includegraphics[width=2.6in]{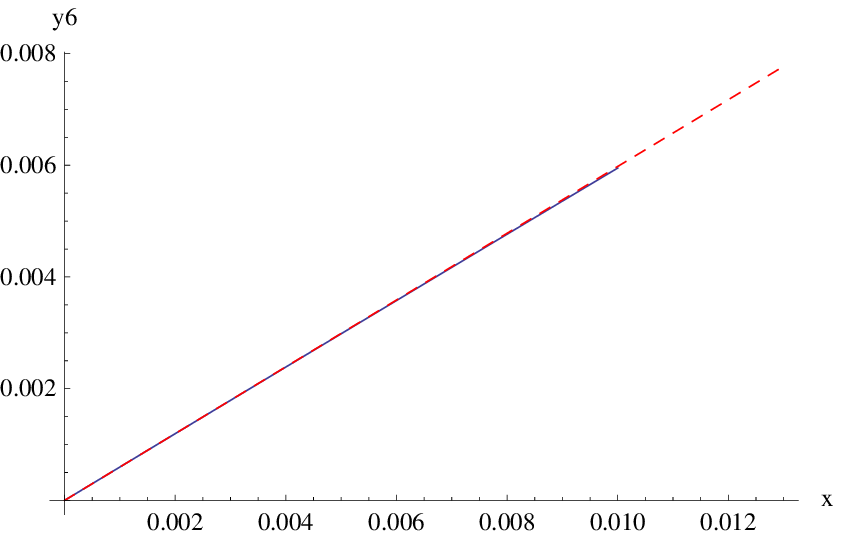}
\end{center}
  \caption{Normalizable coefficients \eqref{fullset} as functions of $\r_{1,1}$. The dashed lines 
represent perturbative predictions \eqref{pertid} with \eqref{table1}.} \label{figure2}
\end{figure}

Using the shooting method of \cite{Aharony:2007vg}, we solve \eqref{vaceoms} and determine 
the normalizable coefficients \eqref{fullset} as a function of $m\ell\equiv (12\r_{1,1})^{1/2}$. 
The results of the computations for small values of $\r_{1,1}$ are collected for numerical test 
in figure  \ref{figure2}. The solid curves are obtained from numerical solution of full 
nonlinear equations \eqref{vaceoms}, and
the dashed lines represent perturbative prediction \eqref{pertid} with \eqref{table1}.

\begin{figure}[t]
\begin{center}
\psfrag{x}{{$m\ell$}}
\psfrag{y}{{$\e$}}
\includegraphics[width=2.6in]{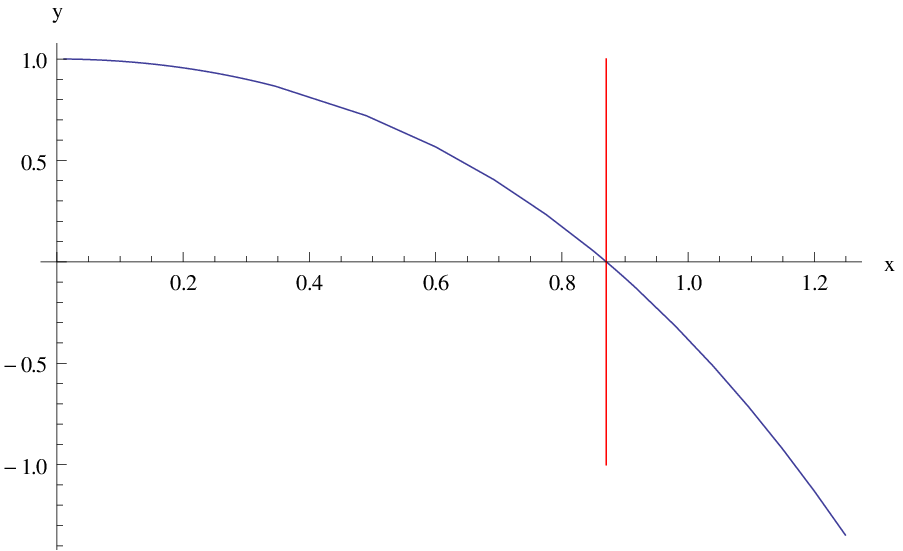}\qquad
\includegraphics[width=2.6in]{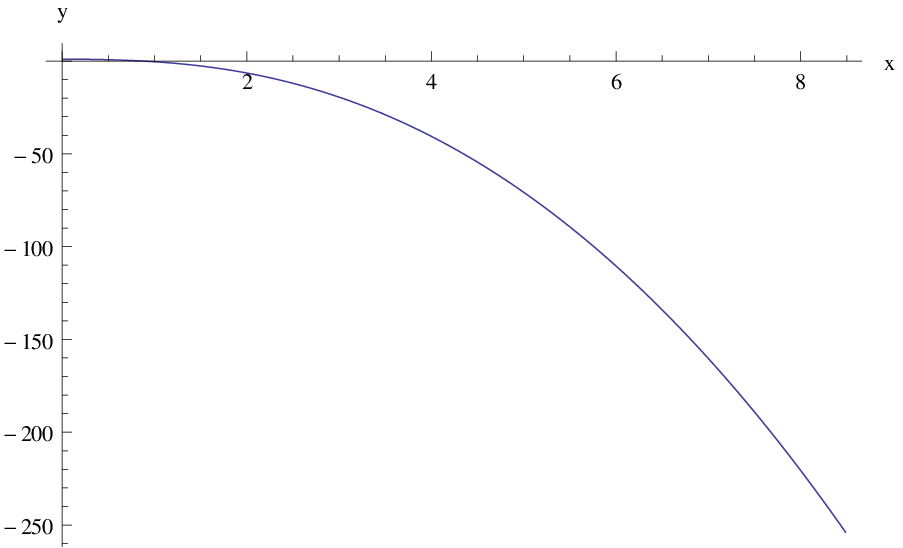}
\end{center}
  \caption{Vacuum energy of the $\caln=2^*$ gauge theory on $S^3$ relative to $\caln=4$ SYM 
Casimir energy, see \eqref{energyfull}. The vertical red line marks vanishing of $\e$, 
see \eqref{m0def}.
 } \label{figure3}
\end{figure}

In full nonlinear numerical analysis we constructed vacua for $0< m\ell \lesssim 8.5 $.
The vacuum energy of the  $\caln=2^*$ gauge theory on $S^3$ relative to $\caln=4$ SYM 
Casimir energy is given by 
\begin{equation}
\begin{split}
\e\equiv \frac{E_{vacuum(m\ell)}}{E_{vacuum}^{\caln=4}}=&
1+\frac{(m\ell)^4}{9}-\frac 23 (1+2\ln 2)(m\ell)^2]\\
&+\biggl\{
32 \r_{1,0}^2+\frac{16}{3}(m\ell)^2 \r_{1,0}+\frac 83 c_{2,0}-4a_{2,0}
\biggr\}\,.
\end{split}
\eqlabel{energyfull}
\end{equation}
It is presented in figure \ref{figure3}. The vertical red line indicates the mass scale 
$m_0\ell$, 
\begin{equation}
\e(m_0\ell)=0\qquad \Longrightarrow\qquad m_0\ell \approx 0.87031\,,
\eqlabel{m0def}
\end{equation}
at which the vacuum energy of the $\caln=2^*$ gauge theory vanishes and becomes negative for even larger value of 
$m\ell$.

\section{Stability of  $\caln=2^*$  vacuum states within BEFP}\label{vacuumstability}

In the previous section we constructed gravitational solutions within PW effective action,
identified as vacua of the $\caln=2^*$ gauge theory on $S^3$. While the complete stability analysis 
of these solutions is beyond the scope of this paper, here we would like to analyze their stability 
within BEFP effective action.  

Effective action describing the fluctuations of an arbitrary PW static solution within 
BEFP has been constructed in \cite{Balasubramanian:2013esa},
\begin{equation}
\begin{split}
&\dd \call\equiv  \call_{BEFP}-\call_{PW}+\calo(X_i^4)\equiv  \dd\call_2+\dd\call_V\,,\\
&\dd\call_2=-(1+c)^2 (\del X_2)^2-\frac {1+c}{4}\left((c^2+c) \r_6^{4/3}-4 (1+c) \r_6^{1/3}
+\frac{4(\del c)^2}{c^2-1}\right) (X_2)^2\,,\\ 
&\dd\call_V=-(1+c)^2 (\del \vec{X}_V)^2-\frac {1+c}{4}\left((c^2-1) \r_6^{4/3}-4 (1+c) \r_6^{1/3}
+\frac{4(\del c)^2}{c^2-1}\right) (\vec{X}_V)^2\,, 
\end{split}
\eqlabel{linear} 
\end{equation}
where $\r_6=\r^6$ and  
$\vec{X}_V=(X_3,X_4,X_5)$ (see section \ref{action} for more details). 
Note that $\dd \call $ is $SU(2)_V$ invariant; as a result it is enough to 
consider a spectrum of only one of $\vec{X}_V$ components. In what follows we choose the latter to be 
$X_3$. 

Introducing 
\begin{equation}
X_2=e^{-i\w t} F_2(y) \Om_s(S^3)\,,\qquad X_3(t,y)=e^{-i\w t} F_3(y)\Om_s(S^3)\,,
\eqlabel{radialeoms}
\end{equation}
where $\Omega_s(S^3)$ are $S^3$ Laplace-Beltrami operator eigenfunctions with eigenvalues $s=l (l+2)$ 
for integer $l$, 
\begin{equation}
\Delta_{S^3}\ \Omega_s(S^3)=-s\ \Omega_s(S^3)=-l (l+2)\ \Omega_s(S^3)\,,
\eqlabel{harmonics}
\end{equation}
we find from \eqref{linear} the following equations of motion
\begin{equation}
\begin{split}
0=&F_2''+F_2' \biggl(\frac{2cc'}{c+1}+\frac{(c^2-1) \r^8}{6a y}
-\frac{4c\r^2}{3a y}+\frac{2 y-1}{y (y-1)}+\frac{1}{a (y-1)}
-\frac{2}{3 a \r^4 y}\biggr)\\
&+\frac{F_2}{4y (1-y) a}  \left(\frac{e^{2 d} \omega^2}{a}-\frac{s}{1-y}\right)
+F_2 \biggl(\frac{(c')^2}{(1-c^2) (c+1)}+\frac{\r^2 (\r^6 c-4)}{4a y^2 (y-1)}\biggr)\,,
\end{split}
\eqlabel{F2eom}
\end{equation}
\begin{equation}
\begin{split}
0=&F_3''+F_3' \biggl(\frac{2cc'}{c+1}+\frac{(c^2-1) \r^8}{6a y}
-\frac{4c\r^2}{3a y}+\frac{2 y-1}{y (y-1)}+\frac{1}{a (y-1)}
-\frac{2}{3 a \r^4 y}\biggr)\\
&+\frac{F_3}{4y (1-y) a}  \left(\frac{e^{2 d} \omega^2}{a}-\frac{s}{1-y}\right)
+F_3 \biggl(\frac{(c')^2}{(1-c^2) (c+1)}+\frac{\r^2 (\r^6 (c-1)-4)}{4a y^2 (y-1)}\biggr)\,.
\end{split}
\eqlabel{F3eom}
\end{equation}
The radial wavefunctions $F_{2,3}$ must be regular at the origin, \ie $z\to 0_+$,
\begin{equation}
F_{2}=z^{l/2}\ f_2^{h} \left( 1+ \calo(z)\right)\,,\qquad F_{3}=z^{l/2}\ f_3^{h} \left( 1+ \calo(z)\right)\,,
\eqlabel{flir}
\end{equation}
and normalizable as $y\to 0_+$,
\begin{equation}
\begin{split}
F_{2}=&y^{3/2}\left(1+y \left(\frac s8-\frac 12 c_{1,0}+\frac{9-\w^2}{8} \right)+\calo(y^2\ln y)\right)\,,\\
F_{3}=&y\left(1+y \left(\frac s4+\frac{4-\w^2}{4} +4 \r_{1,1}-2 \r_{1,0}
-\frac16 c_{1,0}-2 \r_{1,1} \ln y \right)+\calo(y^2\ln y)\right)\,.
\end{split}
\eqlabel{fluv}
\end{equation}
Note that we set the normalizable coefficient of $F_{2,3}$ in the UV to one.

\begin{figure}[t]
\begin{center}
\psfrag{x}{{$m\ell$}}
\psfrag{y}{{$\w_{2,\{n,l\}}/\w_{2,\{n,l\}}^{SYM}$}}
\psfrag{z}{{$\w_{3,\{n,l\}}/\w_{3,\{n,l\}}^{SYM}$}}
\includegraphics[width=2.6in]{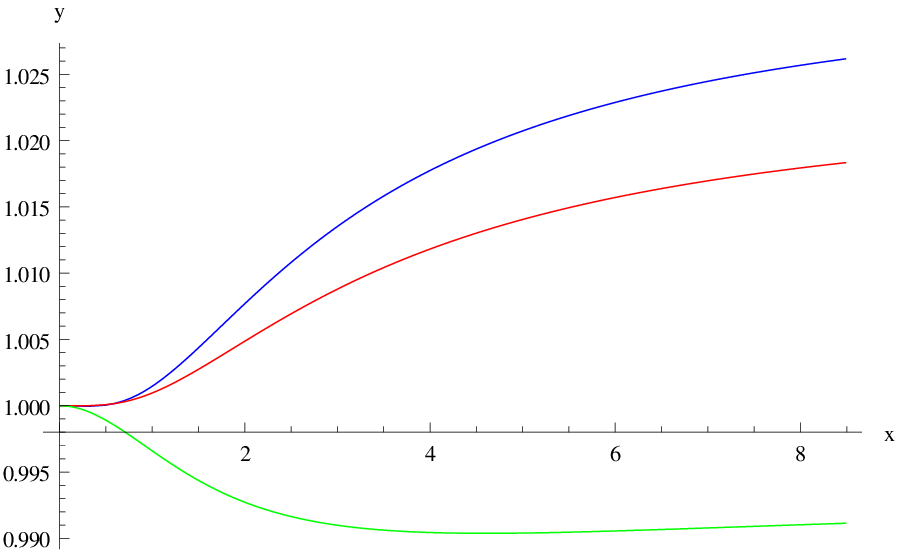}
\qquad
\includegraphics[width=2.6in]{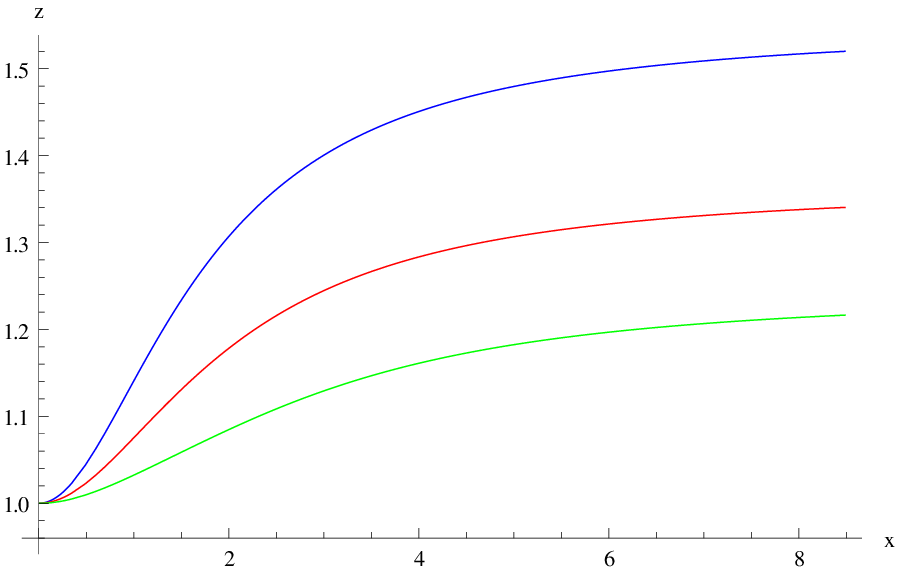}
\end{center}
  \caption{Low energy states in the spectrum of BEFP fluctuations about PW vacua:
 $\{n,l\}=\{(0,0)\,;\, (0,1)\,;\, (1,0) \}$ (blue, red, green). See section \ref{vacuumstability}. 
 } \label{figure4}
\end{figure}

When both scalars of the PW flow are set to zero, \eqref{F2eom}-\eqref{fluv} 
corresponds to fluctuations of gravitational 
modes dual to dimension-3 (for $F_2$) and dimension-2 (for $F_3$) operators of the $\caln=4$ SYM
on $S^3$.  In this case the equations can be solved analytically. 
We find,  
\begin{equation}
\begin{split}
F_{2,\{n,l\}}^{SYM}=&y^{3/2} (1-y)^{l/2}\ _2F_1\biggl(-n\,,3+n+l\,\,; l+2\,\,; 1-y\biggr)\,,\\
\w_{2,\{n,l\}}^{SYM}=&3+2 n +l\,,
\end{split}
\eqlabel{f2solve}  
\end{equation}   
\begin{equation}
\begin{split}
F_{3,\{n,l\}}^{SYM}=&y (1-y)^{l/2}\ _2F_1\biggl(-n\,,2+n+l\,\,; l+2\,\,; 1-y\biggr)\,,\\
\w_{3,\{n,l\}}^{SYM}=&2+2n +l\,,
\end{split}
\eqlabel{f3solve}  
\end{equation}   
where $\{n,l\}$ are non-negative integers. 
For supersymmetric PW flows \eqref{n2susy} we have to resort to numerics. 
The results of the numerical analysis are presented in figure \ref{figure4}.
We look at the states with  $\{n,l\}=\{(0,0)\,;\, (0,1)\,;\, (1,0) \}$
for both $F_2$ and $F_3$ radial functions. Over the range of parameters
discussed, the embedding of PW flows within BEFP effective action is stable.

\section{Black hole spectrum in PW effective action}\label{bh}

We begin with the metric ansatz and the boundary conditions representing regular 
Schwarzschild black hole solutions in PW effective action with the $S^3$ horizon. 
We explain how the normalizable coefficients of the gravitational solution encode
the thermodynamic properties of the black holes: the temperature $T_{BH}$, the energy $E_{BH}$, 
the entropy $S_{BH}$ 
and the free energy $\calf_{BH}$. We define the size $\ell_{BH}$ of a black hole as 
\begin{equation}
\left(\frac{\ell_{BH}}{L}\right)^3\equiv \frac{A_{horizon}}{L^3}\,.
\eqlabel{ahor}
\end{equation}   
We compute excitation energy $\Delta(\ell_{BH}/L\,, (m\ell))$,
\begin{equation}
\Delta(\ell_{BH}/L\,, (m\ell)) = \frac{E_{BH}(\ell_{BH}/L\,, m\ell)-E_{vacuum}(m\ell)}{E_{vacuum}^{\caln=4}}\,,
\eqlabel{defDelta}  
\end{equation}
as a function of $\ell_{BH}/L$, but for select values of $m\ell$:
\nxt perturbatively in $m\ell$, to order $\calo((m\ell)^2)$;
\nxt for $\r_{1,1}=\frac{1}{12}(m\ell)^2=\{1,1.5,2,\cdots 5,5.5,5.8\}$ (the last value 
corresponds to the largest value of $m\ell$ for which we computed $E_{vacuum}$);\\
and present a strong numerical evidence that 
\begin{equation}
\lim_{\ell_{BH}/L\to 0} \Delta(\ell_{BH}/L\,, (m\ell)) = 0\,.
\eqlabel{limigap}
\end{equation}
Thus, we conclude that there is no gap in the spectrum of black holes in PW geometry; 
correspondingly, there is no gap in $SO(4)$-invariant equilibrium states of 
the $\caln=2^*$ gauge theory on $S^3$ in the planar limit 
and for large 't Hooft coupling, as there is no energy gap for generic $SO(4)$-invariant 
excitations in this theory. 

\subsection{Metric ansatz and the boundary conditions for black holes in PW}

Recall that the vacuum solutions of section \ref{vacuum}  
were obtained within metric ansatz \eqref{geomdyn},
\begin{equation}
\begin{split}
ds_5^2\bigg|_{vacuum}=&\frac{4}{\cos^2 x} \left(-a e^{-2d} (dt)^2+\frac{(dx)^2}{a}+\sin^2 x (d\Omega_3)^2\right)\\
=&\frac{4}{y} \left(-a e^{-2d} (dt)^2+\frac{(dy)^2}{4y(1-y)a}+(1-y) (d\Omega_3)^2\right)\,,
\end{split}
\eqlabel{metricbh}
\end{equation} 
where in the second line we recalled the radial coordinate $y=\cos^2 x$, $y\in[0,1]$. Regularity at the 
origin ($y\to 1_-$) required that the metric functions $a$ and $d$ remain finite and non-zero. Notice that the 
three-sphere shrinks to zero size in this limit. 

In close analogy to \eqref{metricbh}, to describe regular horizon black holes, we reparameterize 
the radial coordinate $y\to y_h y$, with a constant $0<y_h<1$, 
while keeping $y\in[0,1]$. We further require that 
$a$ has a simple zero  and $d$ remains finite as $y\to 1_-$:
\begin{equation}
\begin{split}
&ds_5^2\bigg|_{BH}
=\frac{4}{y_h y} \left(-a e^{-2d} (dt)^2+\frac{y_h (dy)^2}{4y(1-y y_h)a}+(1-y y_h) (d\Omega_3)^2\right)\,,\\
&0<y_h<1\,,\qquad y\in[0,1]\,,\qquad \lim_{y\to 1_-} a=0\,,\\
&\lim_{y\to 1_-} a'={\rm finite}\ne 0\,,\qquad \lim_{y\to 1_-} d={\rm finite}\,.
\end{split}
\eqlabel{s5bh}
\end{equation}
Given \eqref{s5bh}, 
\begin{equation}
A_{horizon}=16\pi^2\frac{(1-y_h)^{3/2}}{y_h^{3/2}}\qquad \Longrightarrow\qquad 
\frac{\ell_{BH}}{L}\equiv\frac{A_{horizon}^{1/3}}{L}=(2\pi^2)^{1/3}\frac{(1-y_h)^{1/2}}{y_h^{1/2}}\,.
\eqlabel{lbh}
\end{equation}
The equations of motion describing black holes \eqref{s5bh} can be obtained from 
\eqref{vaceoms} with the simple change of variables\footnote{We used the last two equations to 
algebraically eliminate $a'$ and $d'$ from the first two.}  $y\to y y_h$,
\begin{equation}
\begin{split}
&0=c''-\frac{c (c')^2}{c^2-1}+c'\biggl(\frac{(c^2-1) \r^8}{6a y}-\frac{4c \r^2}{3a y}+\frac{a (2 y y_h-1)+y y_h}{y a (y y_h-1)}
-\frac{2}{3 y a \r^4}\biggr)
\\
&- \frac{\r^2 (c^2-1) (\r^6 c-4)}{4(1-yy_h) y^2 a}\,,\\
&0=\r''-\frac{(\r')^2}{\r}+\r' \biggl(\frac{(c^2-1) \r^8}{6a y}-\frac{4c \r^2}{3a y}+\frac{a (2 y y_h-1)+y y_h}{y a (y y_h-1)}
-\frac{2}{3 y a \r^4}\biggr)\\
&
-\frac{(c^2-1) \r^9}{12(1-yy_h) y^2 a}-\frac{1-\r^6 c}{6\r^3 y^2 a (1-yy_h)}\,,\\
&0=d'-\frac{2 y (1-yy_h) (c')^2}{3(c^2-1)}-\frac{8 (1-yy_h) y (\r')^2}{\r^2}\,,\\
0&=a'-(y-y^2y_h)a\left(\frac{8    (\r')^2}{\r^2}+\frac{2  (c')^2}{3(c^2-1)}\right)+\frac{(yy_h-2) a+yy_h}{y (1-yy_h)}
-\frac{(c^2-1) \r^8-8\r^2 c}{6y}\\
&+\frac{2}{3 y \r^4}\,.
\end{split}
\eqlabel{bheoms}
\end{equation}  
The boundary conditions in the UV, \ie $y\to 0_+$,  specify the asymptotic
expansion   
\begin{equation}
\begin{split}
a=&1+y\ \frac 23  \hc_{1,0}\  + y^2\ \biggl(\ha_{2,0}+\biggl(\frac 23 \hc_{1,0}(\hc_{1,0}+y_h)+8\hr_{1,1}^2+
16\hr_{1,1}\hr_{1,0}\biggr)\ln y\\
&+8 \hr_{1,1}^2\ln^2 y\biggr)+ \calo(y^3\ln^3 y)\,,\\
d=&y\ \frac13  \hc_{1,0}+y^2\ \biggl(\frac12 \hc_{2,0}-\frac{1}{36} \hc_{1,0}^2+4 \hr_{1,0}^2-\frac18 \hc_{1,0} y_h+2 \hr_{1,1}^2
+4 \hr_{1,0} \hr_{1,1}\\
&+\biggl(\frac 14 \hc_{1,0} y_h+\frac13 \hc_{1,0}^2+4 \hr_{1,1}^2+8 \hr_{1,0} \hr_{1,1}\biggr) \ln y
+4 \hr_{1,1}^2 \ln^2 y\biggr)+ \calo(y^3\ln^3 y)\,,\\
\r=&1+y\ \left(\hr_{1,0}+\hr_{1,1} \ln y\right)+y^2\ 
\biggl(
\frac{1}{12} \hc_{1,0}^2+\hr_{1,0}y_h-3 \hr_{1,1} \hc_{1,0}+6 \hr_{1,1}^2\\
&-4 \hr_{1,0} \hr_{1,1}+\frac43 \hc_{1,0} \hr_{1,0}+\frac32 \hr_{1,0}^2
+\biggl(\frac43 \hr_{1,1} \hc_{1,0}+\hr_{1,1} y_h-4 \hr_{1,1}^2\\
&+3 \hr_{1,0} \hr_{1,1}\biggr) \ln y+\frac32 \hr_{1,1}^2 \ln^2 y
\biggr)+\calo(y^3\ln^3 y)\,,\\
c=&1+y\ \hc_{1,0}+y^2\ \biggl(\hc_{2,0}+\biggl(\frac12 \hc_{1,0}y_h+\frac23 \hc_{1,0}^2\biggr) \ln y\biggr)+ \calo(y^3\ln^2 y)\,.
\end{split}
\eqlabel{bhuv}
\end{equation}
In \eqref{bhuv} the non-normalizable coefficients $\hr_{1,1}$ and $\hc_{1,0}$ are related to 
corresponding coefficients of the vacuum solution as 
\begin{equation}
\hr_{1,1}=y_h \r_{1,1}\,,\qquad \hc_{1,0}=y_h c_{1,0}\,,
\eqlabel{bhmatch}
\end{equation}  
to be further matched with the mass parameters $\{m_b,m_f\}$ of the dual gauge theory as in \eqref{genflow}.  
The rest of the  coefficients in \eqref{bhuv} are normalizable. The asymptotic expansion in the IR,
\ie as $z=(1-y)\to 0_+$ is different from the one in \eqref{originz} --- here it reflects the presence of a 
regular horizon (see \eqref{s5bh}),   
\begin{equation}
\begin{split}
&a=\frac z6 \left(\left(1-(\hc^h_0)^2\right)(\hr_0^h)^8+8\hc^h_0(\hr^h_0)^2+\frac{4}{(\hr^h_0)^4}+\frac{6y_h}{1-y_h}\right)
+\calo(z^2)\,,\\
&d=\hd^h_0+\calo(z)\,,\\
&\r=\hr^h_0+\calo(z)\,,\\
&c=\hc^h_0+\calo(z)\,.\\
\end{split}
\eqlabel{bhir}
\end{equation}
The full set of the non-normalizable coefficients is 
\begin{equation}
\{\ha_{2,0}\,,\ \hr_{1,0}\,,\ \hc_{2,0}\,,\ \hr^h_0\,,\ \hc^h_0\,,\ \hd^h_0\}\,.
\eqlabel{bhfullset}
\end{equation} 
Note that we have the correct number of non-normalizable coefficients to uniquely specify a solution of 
two second-order and two first-order ODEs given a choice of \eqref{bhmatch}.

\subsubsection{Perturbative black holes solutions}
As in section \ref{vsmall}, we can construct solutions to \eqref{bheoms}-\eqref{bhir} 
perturbatively in $m\ell$ to order $\calo((m\ell)^2)$. 

We introduce 
\begin{equation}
\begin{split}
&c=\cosh(2\l \hat{\chi}_1(y)+\calo(\l^3))\,,\qquad \r=e^{\l^2 \hat\a_2(y)+\calo(\l^4)}\,,\\
&a=\frac{(1-y)(1+y(1-y_h))}{1-yy_h}+\l^2 \hat a_2(y)+\calo(\l^4)\,,\qquad d=\l^2 \hat d_2(y)+\calo(\l^2)\,,
\end{split}
\eqlabel{pertbh}
\end{equation}
where $\l$ is a small parameter.  Substituting \eqref{pertvacuum} into \eqref{vaceoms} we find
\begin{equation}
\begin{split}
&0=\hat\c_1''-\frac{\hat\c_1'}{y(1-y)}\biggl(1+y+\frac{y(1-y_h)((2-y)y y_h-2)}{(1-yy_h)(1+y(1-y_h))}\biggr)+\frac {3\hat\c_1}{4y^2(1-y)(1+y(1-y_h))}\,,\\
&0=\hat\a_2''-\frac{\hat\a_2'}{y(1-y)}\biggl(1+y+\frac{y(1-y_h)((2-y)y y_h-2)}{(1-yy_h)(1+y(1-y_h))}\biggr)+\frac {\hat\a_2}{y^2(1-y)(1+y(1-y_h))}\,,\\
&0=\hat a_2'-\frac{2-yy_h}{y(1-yy_h)}\hat a_2-\frac 83 y(1-y)(1+y(1-y_h))(\hat\c_1')^2+\frac 2y(\hat\c_1)^2\,,\\
&0=\hat d_2'-\frac 83 y(1-y y_h)(\hat\c_1')^2\,.
\end{split}
\eqlabel{perbh}
\end{equation}
For the asymptotic expansions we have:
\nxt as $y\to 0_+$, 
\begin{equation}
\begin{split}
\hat\c_1=&y^{1/2} \left(1+y\ \left(\hat\c_{1,0,(1)} +\frac {y_h}{4} \ln y\right)+\calo(y^2\ln y)\right)\,,\\
\hat\a_2=&\hr_{1,1,(2)}\biggl(\left(\hat\a_{1,0,(2)}+\ln y\right)y+\calo(y^2\ln y)\biggr)\,,\\
\hat a_2=&\frac 43 y + y^2\ \left( \hat a_{2,0,(2)}+\frac {4y_h}{3}\ln y\right)+\calo(y^3\ln^2 y)\,,\\
\hat d_2=&\frac 23 y +y^2\ \left(-\frac {y_h}{4} +2 \hat\c_{1,0,(1)} +\frac {y_h}{2} \ln y\right)
+\calo(y^3\ln^2 y)\,,
\end{split}
\eqlabel{pertuvbh}
\end{equation}
\nxt as $z\to 0_+$
\begin{equation}
\begin{split}
\hat\c_1=&\hat\c_{0,(1)}^{h}\left(1-\frac {3}{4(2-y_h)} z+\calo(z^2)\right)\,,\\
\hat\a_2=&\hr_{1,1,(2)}\biggl(\hat\a_{0,(2)}^{h}\left(1-\frac {1}{(2-y_h)} z+\calo(z^2)\right)\biggr)\,,\\
\hat a_2=&2(\hat\c_{0,(1)}^{h})^2 z +\calo(z^2)\,,\\
\hat d_2=&\hat d^h_{0,(2)}-\frac{3(1-y_h)}{2(2-y_h)^2}(\hat\c_{0,(1)}^{h})^2 z +\calo(z^2)\,.
\end{split}
\eqlabel{pertirbh}
\end{equation}\\
Equations \eqref{perbh}-\eqref{pertuvbh} have to be solved numerically for different values of $y_h$.
 
To compare with the full numerical solution, we identify, to order $\calo(\l^2)$, 
\begin{equation}
\begin{split}
&\hr_{1,1}=\hr_{1,1,(2)} \l^2\,,\qquad \hc_{1,0}=2 \l^2\,,\qquad \hr_{1,0}=\hr_{1,1,(2)}\hat\a_{1,0,(2)} \l^2\,,\qquad 
\hc_{2,0}=4\hat\c_{1,0,(1)} \l^2\,, \\
&\ha_{2,0}=y_h-1+\hat a_{2,0,(2)}\l^2\,,\qquad \hr^h_0=1+\hr_{1,1,(2)}\hat\a_{0,(2)}^{h}\l^2\,,\\
&\hc^h_0=1+2(\hat\c_{0,(1)}^{h})^2\l^2\,,\qquad 
\hd^h_0=\hat d^h_{0,(2)}\l^2\,.
\end{split}
\eqlabel{pertidbh}
\end{equation}
Note that $\caln=2$ supersymmetry in the UV at $\calo(\l^2)$ leads to (see \eqref{n2susy})
\begin{equation}
\hr_{1,1,(2)}=\frac 13 \,.
\eqlabel{r112bh}
\end{equation}

\subsection{Thermodynamic properties of black holes in PW}

Requiring that there is no conical singularity in the analytical continuation 
$t\to i t_E$ of the metric \eqref{s5bh}  as $y\to 1_-$ we compute the
Hawking temperature $T_{BH}$ of the black hole using \eqref{bhir},
\begin{equation}
\begin{split}
T_{BH}=&\frac{e^{-\hd^h_0}}{12 \pi y_h^{1/2}(1-y_h)^{1/2}} \biggl((1-y_h) (1-(\hc^h_0)^2) 
(\hr^h_0)^8
+8 \hc^h_0 (1-y_h) (\hr^h_0)^2+6 y_h\\
&+\frac{4 (1-y_h)}{(\hr^h_0)^4}\biggr)\,.
\end{split}
\eqlabel{tbh}
\end{equation}
The Bekenstein-Hawking entropy of the black hole is given by 
\begin{equation}
S_{BH} = \frac{A_{horizon}}{4 G_5}=\frac{4\pi^2}{G_5}\ \frac{(1-y_h)^{3/2}}{y_h^{3/2}}\,.
\eqlabel{sbh}
\end{equation}
The free energy $\calf_{BH}$ can be computed following holographic 
renormalization procedure discussed in section \ref{holren}.
We find
\begin{equation}
\begin{split}
&\calf_{BH}=\frac{3\pi}{4 G_5}\ 
\biggl(1+\frac{\hc_{1,0}^2}{y_h^2} \left(\frac 49-\frac{16}{9}\ln 2
+\frac 89\ln y_h\right)
+\frac{\hc_{1,0}}{y_h}\left(-\frac 43-\frac83 \ln 2+\frac 43 \ln y_h\right)
\\
&+ 32 \frac{\hr_{1,1}^2}{y_h^2}\left(2\ln 2-\ln y_h\right)
+\frac{1}{y_h^2}\biggl\{64 \hr_{1,1}\hr_{1,0}+\frac 83 \hc_{2,0}+32 \hr_{1,0}^2
-4 \ha_{2,0}\biggr\}
\biggr)\\
&-\frac{(1-y_h)\pi e^{-\hd^h_0}}{3y_h^2 G_5}\biggl((1-y_h) (1-(\hc^h_0)^2) 
(\hr^h_0)^8
+8 \hc^h_0 (1-y_h) (\hr^h_0)^2+6 y_h
+\frac{4 (1-y_h)}{(\hr^h_0)^4}\biggr)\,.
\end{split}
\eqlabel{fbh}
\end{equation} 
The contribution in the last line in \eqref{fbh} comes from the lower limit of integration of the bulk 
contribution to the regularized free energy, \eqref{totder}; it equals precisely 
to $(-S_{BH}T_{BH})$. Computing the holographic stress-energy tensor, as described in \cite{Buchel:2004hw}
we find 
\begin{equation}
\begin{split}
E_{BH}=&\frac{3\pi}{4 G_5}\ 
\biggl(1+\frac{\hc_{1,0}^2}{y_h^2} \left(\frac 49-\frac{16}{9}\ln 2
+\frac 89\ln y_h\right)
+\frac{\hc_{1,0}}{y_h}\left(-\frac 43-\frac83 \ln 2+\frac 43 \ln y_h\right)
\\
&+ 32 \frac{\hr_{1,1}^2}{y_h^2}\left(2\ln 2-\ln y_h\right)
+\frac{1}{y_h^2}\biggl\{64 \hr_{1,1}\hr_{1,0}+\frac 83 \hc_{2,0}+32 \hr_{1,0}^2
-4 \ha_{2,0}\biggr\}
\biggr)\\
=&\frac{3N^2}{16\ell}\biggl(1+\frac{(m\ell)^4}{9}-\frac 23 (1+2\ln 2-\ln y_h)(m\ell)^2
\\&+\frac{1}{y_h^2}\biggl\{
32 \hr_{1,0}^2+\frac{16}{3}(m\ell)^2 y_h\hr_{1,0}+\frac 83 \hc_{2,0}-4\ha_{2,0}
\biggr\}
\biggr)\,,
\end{split}
\eqlabel{ebh}
\end{equation}
where in the last line we expressed the energy in terms of the dual gauge theory variables 
using \eqref{bhmatch} and \eqref{matchPW}. Notice that the basic thermodynamic relation,
\begin{equation}
\calf_{BH}=E_{BH}-S_{BH}T_{BH}\,,
\eqlabel{brelation}
\end{equation}
is satisfied automatically. 

Using \eqref{pertidbh}, from \eqref{ebh} we have 
\begin{equation}
\begin{split}
\frac{E_{BH}}{E_{vacuum}^{\caln=4}}=&1+\frac{4(1-y_h)}{y_h^2}+\left(\frac{8}{3y_h}\hat\c_{1,0,(1)}-\frac{1}{y_h}\hat a_{2,0,(2)}
-\frac 23(1+2\ln 2-\ln y_h)\right)(m\ell)^2\\
&+\calo((m\ell)^4)\,.
\end{split}
\eqlabel{epbh}
\end{equation}

\subsection{$ \Delta(\ell_{BH}/L\,, (m\ell))$}  
We are now ready to present results for $ \Delta(\ell_{BH}/L\,, (m\ell))$ as defined by \eqref{defDelta}.

\begin{figure}[t]
\begin{center}
\psfrag{x}{{$\ell_{BH}/L$}}
\psfrag{y}{{$\Delta_2$}}
\includegraphics[width=4in]{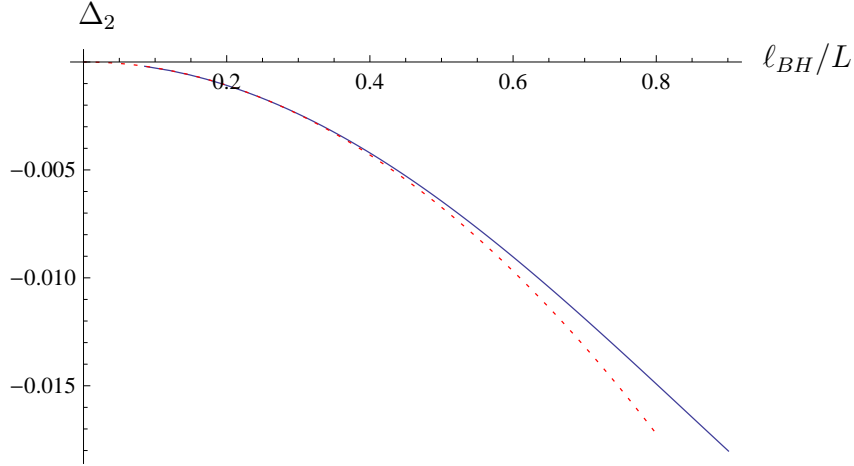}
\end{center}
  \caption{Solid  line represents $\Delta_2$ as defined in \eqref{perDelta}. The dotted red line 
represents the best quadratic fit to the first 10$\%$ of data points, see \eqref{delta2fit}.} \label{figure5}
\end{figure}

\begin{figure}[t]
\begin{center}
\psfrag{x}{{$\ell_{BH}/L$}}
\psfrag{y}{{$\Delta$}}
\psfrag{z}{{$\Delta(m\ell=8.34266)$}}
\includegraphics[width=2.6in]{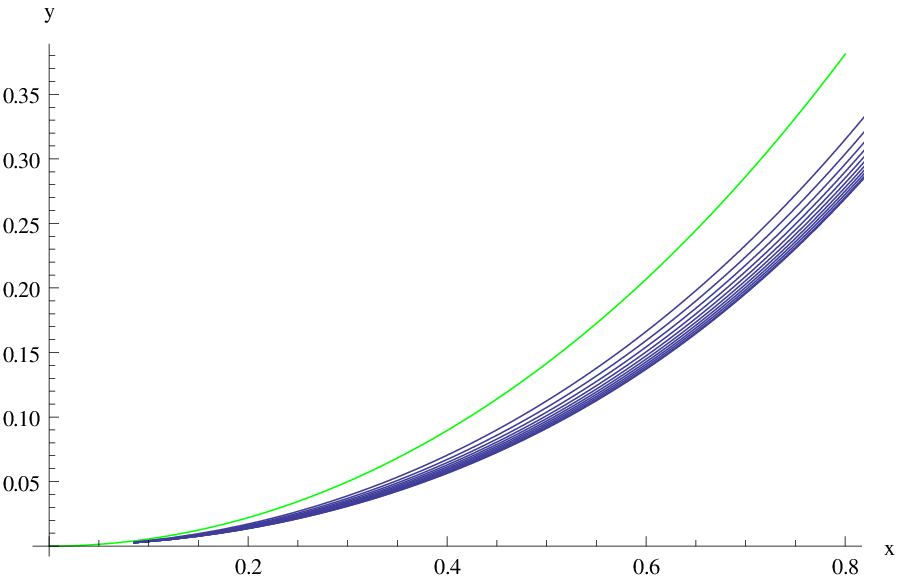}\qquad
\includegraphics[width=2.6in]{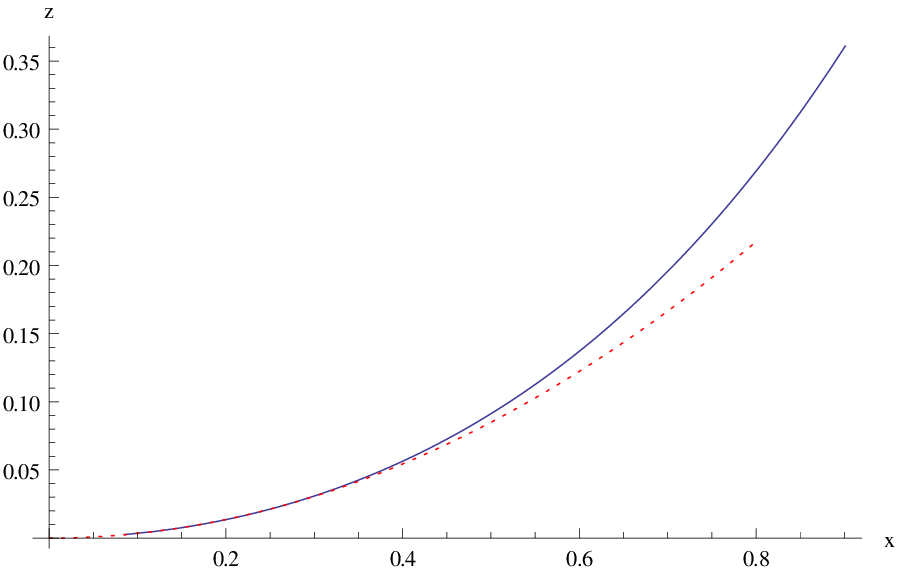}
\end{center}
  \caption{Left panel: Black hole mass gap relative to $E_{vacuum}^{\caln=4}$, see \eqref{defDelta},
as a function of $\ell_{BH}/L$ for select 
values of $m\ell$. The green curve represents 
$\Delta(m\ell=0)$.  Right panel: $\Delta$ for the largest value of $m\ell$ computed, 
$m\ell=8.34266$; the dotted red line represents 
the best quadratic fit to the first 10$\%$ of data points, see \eqref{deltafit}.} \label{figure6}
\end{figure}

To order $\calo((m\ell)^2)$, using \eqref{epvacuum} and \eqref{epbh}, we find
\begin{equation}
\begin{split}
\Delta=&\frac{4(1-y_h)}{y_h^2}+\Delta_2\ (m\ell)^2+\calo((m\ell)^4)\,,\\
\Delta_2=&\Delta_2(y_h)=\frac83\left(\frac{\hat\c_{1,0,(1)}}{y_h}-\c_{1,0,(1)}\right)-\left(\frac{\hat a_{2,0,(2)}}{y_h}-a_{2,0,(2)}\right)+\frac 23 \ln y_h\,.
\end{split}
\eqlabel{perDelta}
\end{equation}
Results of numerical computations of $\Delta_2$ are presented in figure \ref{figure5}. A solid line  
represents the data points, and the red dotted line is the best quadratic fit using the first 10$\%$ of data points:
\begin{equation}
\Delta_2\bigg|_{fit}=-0.0269118 \biggl(\frac{\ell_{BH}}{L}\biggr)^2\,.
\eqlabel{delta2fit}
\end{equation}  
Our numerical results present a strong evidence that 
\begin{equation}
\lim_{\ell_{BH}/L\to 0} \Delta_2=0\,,
\eqlabel{limpert}
\end{equation}
as a result, we see that $\Delta$ vanishes in this limit to order $\calo((m\ell)^2)$.

Using \eqref{freeenergy} and \eqref{ebh} we compute $\Delta$ for 
$\r_{1,1}=\frac{1}{12}(m\ell)^2=\{1,1.5,2,\cdots 5,5.5,5.8\}$. 
The results are presented in the left panel of figure \ref{figure6}
(the top-to-bottom blue curves correspond to $\r_{1,1}$ variation $1\to 5.8$). 
The green curve represents 
$\Delta(m\ell=0)$:
\begin{equation}
\Delta(m\ell=0)=\frac{2^{4/3}}{\pi^{4/3}}\ \biggl(\frac{\ell_{BH}}{L}\biggr)^2
+\frac{2^{2/3}}{\pi^{8/3}}\ \biggl(\frac{\ell_{BH}}{L}\biggr)^4\,.
\eqlabel{deltan4}
\end{equation}
The right panel represents $\Delta$ for the largest value of $m\ell$ computed:  $m\ell=8.34266$, 
with the red dotted 
line indicating the best quadratic fit to the first $10\%$ of data points:
\begin{equation}
\Delta(m\ell=8.34266)\bigg|_{fit}=0.339765 \biggl(\frac{\ell_{BH}}{L}\biggr)^2\,.
\eqlabel{deltafit}
\end{equation}  
 Note that for $m\ell=8.34266$,  $\e=-243.785$, implying that for 
the smallest size black hole studied, $\ell_{BH}/L=0.0855056$, 
\begin{equation}
\frac{E_{BH}-E_{vacuum}}{E_{vacuum}}=1.04285\times 10^{-5}\,.
\eqlabel{smallest}
\end{equation}
We conclude that numerical results strongly  suggest \eqref{limigap}.

~\\
\section*{Acknowledgments}
I would like to thank Colin Denniston, Martin Kruczenski, Luis Lehner, Steve Liebling  and Volodya Miransky
for valuable discussions.  
I thank the Galileo Galilei Institute for Theoretical Physics for the hospitality and the INFN for 
partial support during the completion of this work.
Research at Perimeter
Institute is supported by the Government of Canada through Industry
Canada and by the Province of Ontario through the Ministry of
Research \& Innovation. I gratefully acknowledge further support by the
NSERC Discovery grant.

\end{document}